\documentclass[aps,pre,twocolumn,superscriptaddress,showpacs,showkeys ]{revtex4-1}

\usepackage{graphicx}                   
\usepackage{hyperref}                   
\usepackage{amsmath,amssymb}            
\usepackage{epstopdf}
\usepackage{subcaption}					

\usepackage{color}
\definecolor{orange}{rgb}{1,0.5,0}
\definecolor{brown}{rgb}{0.65, 0.16, 0.16}
\definecolor{phlox}{rgb}{0.87, 0.0, 1.0}

\begin{document}

\title{Flicker Noise in Two-Dimensional Electron Gas}

\author{M. N. Najafi}
\affiliation{Department of Physics, University of Mohaghegh Ardabili, P.O. Box 179, Ardabil, Iran}
\email{morteza.nattagh@gmail.com}

\author{S. Tizdast}
\affiliation{Department of Physics, University of Mohaghegh Ardabili, P.O. Box 179, Ardabil, Iran}
\email{Susan.tizdast@gmail.com}

\author{Z. Moghaddam}
\affiliation{Department of Physics, University of Mohaghegh Ardabili, P.O. Box 179, Ardabil, Iran}
\email{zahramoghadam.physics@gmail.com }

\author{M. Samadpour}
\affiliation{Physics Department, K. N. Toosi University of Technology, Tehran, Iran.}
\email{samadpour@kntu.ac.ir}

\begin{abstract}
Using the method developed in a recent paper (Euro. Phys. J. B 92.8 (2019): 1-28) we consider $1/f$ noise in two-dimensional electron gas (2DEG). The electron coherence length of the system is considered as a basic parameter for discretizing the space, inside which the dynamics of electrons is described by quantum mechanics, while for length scales much larger than it the dynamics is semi-classical. For our model, which is based on the Thomas-Fermi-Dirac approximation, there are two control parameters: temperature $T$ and the disorder strength ($\Delta$). Our Monte Carlo studies show that the system exhibits $1/f$ noise related to the electronic avalanche size, which can serve as a model for describing the experimentally observed flicker noise in 2DEG. The power spectrum of our model scales with frequency with an exponent in the interval $0.3<\alpha_{PS}<0.6$. We numerically show that the electronic avalanches are scale invariant with power-law behaviors in and out of the metal-insulator transition line.
\end{abstract}


\maketitle

\section{Introduction}
Two-dimensional electron gas (2DEG) has been a challenging topic in condensed matter physics for many years. The challenging behaviors of 2DEG, like the metal-insulator transition (MIT) and $1/f$ noise, have put it at the center of a long-running debates~\cite{abrahams2001metall,kravchenko2003meta,najafi2019electronic,weissman19881}. According to one-parameter scaling theory~\cite{abrahams2001metall}, in the absence of spin interaction and scattering no MIT is expected in 2D, the fact that was violated by the experiments of Kravchenko~\cite{kravchenko1995scaling}. Despite of the huge attempts for explanation of this MIT ~\cite{abrahams1979scaling,dolan1979nonmetallic}, it has been remaining as a big mystery in the condensed matter physics, for a good review see~\cite{najafi2019electronic}. $\frac{1}{f}$ (flicker) noise in 2DEG is another big mystery low-dimensional condensed matter whose source is not known yet. In nature the flicker noise is referred to the systems in which the power spectrum of a time signal $S(t)$ (event sizes or other observables) shows power-law decay and heavy tails at low frequencies (in the form $ \frac{1}{f^\alpha } $) which is attributed to the scalelessness of the dynamics of the system. Various experiments reported on $1/f$ noise for some observables like the electronic resistance of the 2DEG $R(t)$, recorded for a fixed temperature and a given disorder. The examples are observed $1/f$ noise in vacuum tube~\cite{johnson1925schottky}, and in carbon composition and thick film resistors~\cite{barry2014measurement2}. The range of exponents that have been reported is $\alpha\in [1.1-1.5]$~\cite{weissman19881,kravchenko2003meta}. The interesting point is that the flicker noise is formed in a \textit{self-organized} fashion, without fine tuning of any parameters, which is a well-known scenario in a large class of \textit{out-of-equilibrium} system, namely self-organized critical (SOC) systems, in which a separation of time scales is needed~\cite{Dhar2006Theoretical}. In the case this scenario is applicable for 2DEG, a big question arises: what is the self-organization mechanism behind this phenomena in 2DEG where the system is in quasi-equilibrium state ("quasi" is due to the fact that the transport in the electronic system is quasi equilibrium) Such a SOC analysis of these systems helps much in understanding the Flicker noise in 2DEG. Much studies have focused on the problem to explain the observed $1/f$ noise in this system, most of which are based on a phenomenological level and have rarely been successful in explaining the phenomena. The example is~\cite{parkhutik2000flicker}, in which a phenomenological approach based on spectroscopy was taken to show that this phenomenon have roots in polarity of the voltage applied to the structure and the degree of dynamic memory. One of the methods used to explain the problem is the Dutta Horn, that we review shortly here. The emperical Hooge's formula~\cite{hooge19761} that was videly used at the time of presentation due to its universality and simplicity~\cite{hooge1981experimental} (stating a simple inverse proportionality between the power spectrum and frequency), was shown by Dutta-Horn that is not universal enough to explain all the range of the exponents that was observed~\cite{weissman19881}. Another main contribution was due to Kogan \textit{et. al.}~\cite{kogan1985low}, in which the essential questions related to semiconductors were explored. Later on some connections with multistep kinetic was explored~\cite{nelkin1983chaos}. Each of these works has its own strengths and weaknesses. For example, Press~\cite{press1978flicker} predicted the importance of higher statistics in $1/f$ noise analysis but erroneously stated that filtered Gaussian noise must depend on properties other than spectrum. We will briefly explore the attempts for explaining this phenomena in this paper. Despite of many studies, there is no well-accepted theory to explain the flicker noise in electron gas systems, and our knowledge is limited to very restricted theoretical studies that can partially explain an aspect of the phenomena. \\
Here we add another possibility to the list, which is based on the recent study on 2DEG which concentrates on the electronic coherence length (CL)~\cite{najafi2019electronic}.  The ingredients of this approach is discretizing the system (in the presence of charged disorder, and zero-interaction limit) according to temperature dependent CL and using the Thomas-Fermi-Dirac approximation for the quantum system for length scales much smaller than CL, and semi-classical Metropolis approach for length scale much larger than CL. The system is assumed to be open, meaning that electrons are permissible to enter/leave the system via leads or the electronic reservoirs, after which the system fastly equilibrates. This approach is able to successfully explain some essential properties of 2DEG MIT, like the universality for weak disorder limit~\cite{kravchenko2003meta,clarke2008impact,altshuler1985electron,shashkin2001indication,vitkalov2001scaling,pudalov1999weak, punnoose2001dilute, prus2001cooling}. It was suggested that (~\cite{najafi2018percolation}) MIT in 2D can be a sole percolation source, in which the system undergoes a percolation transition in a line in the temperature-disorder ($T-\Delta$) phase diagram at which the electronic comprehensibility diverges. In this model the coherence length was taken into account as the main building block of the electronic dynamics, which separates the quantum dynamics from the semi-classical one. Using this approach we show that $\frac{1}{f}$ noise is seen over the parameter space. We show that $1/f$ noise arises in this system.
Since in this article we examine the relationship between $1 / f $ noise and 2D electron gas, for a better study, we specifically explain this noise in condensed matter and explicate some of the theories presented.\\
The article is arranged as follows: In the following subsection, we introduce $1/f$ noise, In the second subsection, we introduce the 2D electron gas as a review of it Then we generalize it in the second section introducing our simulation in this paper. In the third section, We present the results obtained. Finally, the article ends with a conclusion.

\section{Flicker noise in condensed matter}
\begin{figure*}
	\begin{subfigure}{0.33\textwidth}\includegraphics[width=\textwidth]{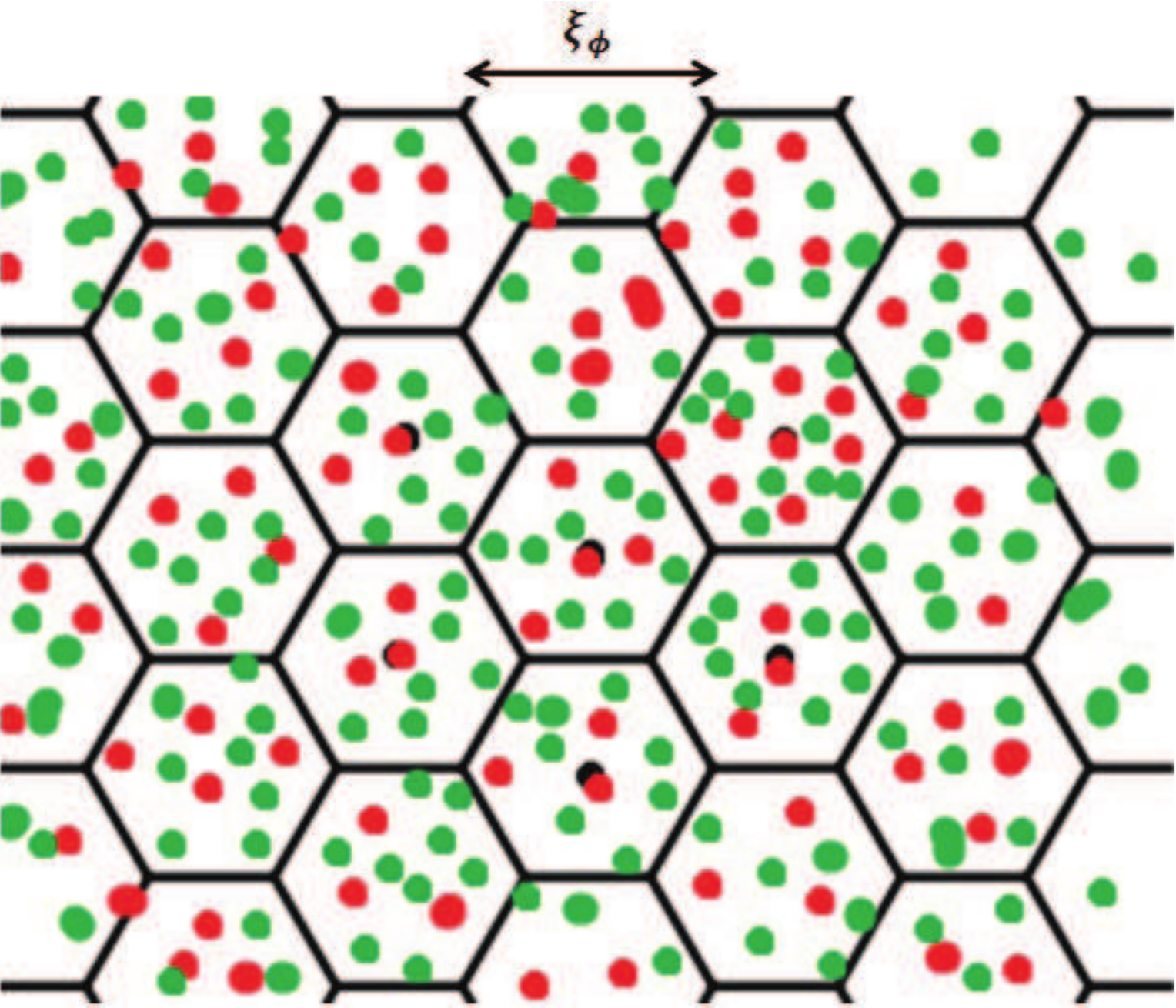}
		\caption{}
		\label{fig:sample1}
	\end{subfigure}
	\begin{subfigure}{0.33\textwidth}\includegraphics[width=\textwidth]{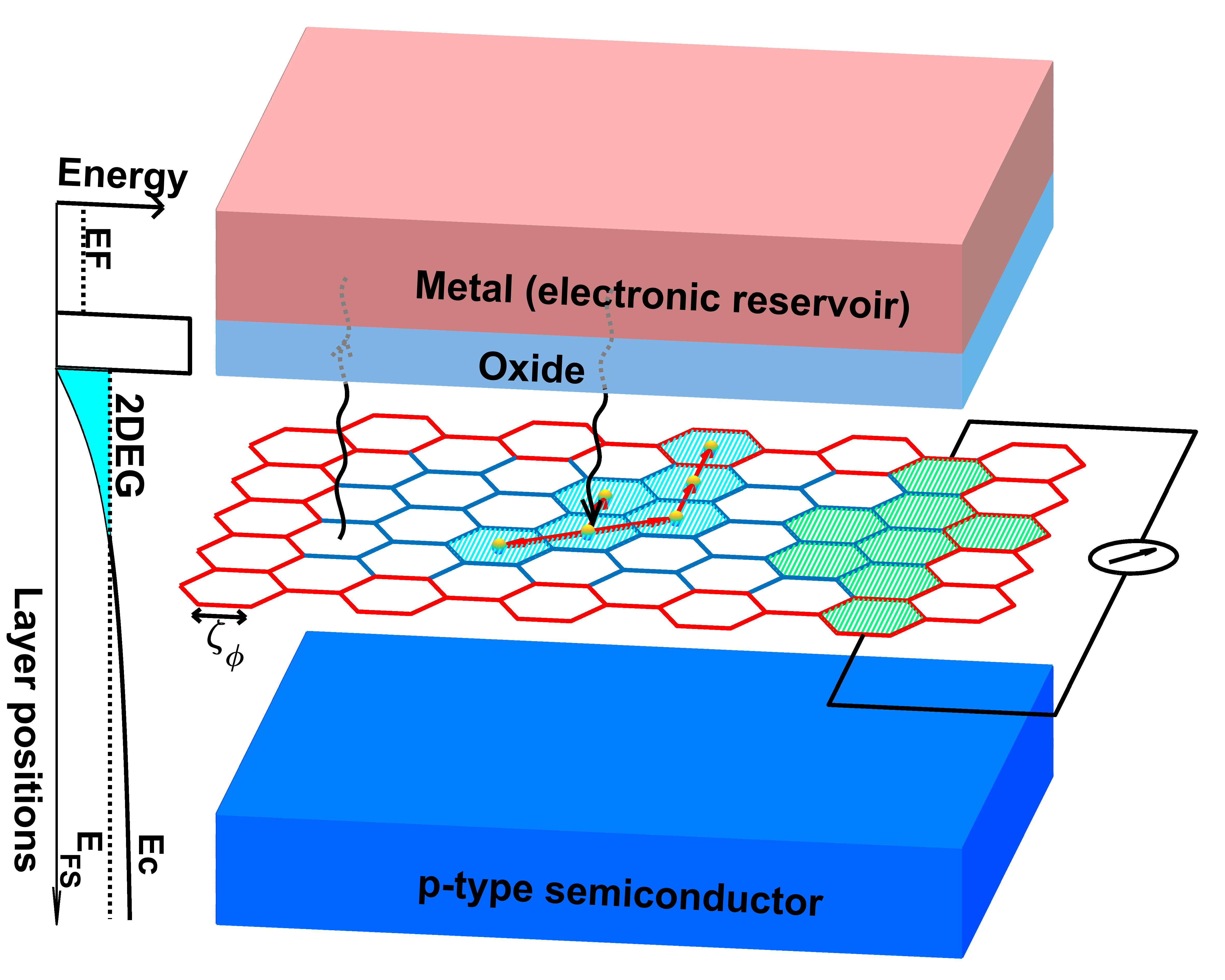}
		\caption{}
		\label{fig:Honeycombb}
	\end{subfigure}
	\caption{(Color Online) (a) A schematic graph of dividing the 2D lattice into many hexagons. Inside the hexagons we have a pure quantum electron gas. The transfer between the cells occurs semi-classically. The green points show electrons and the red ones show the impurities. (b) A schematic set up of a 2D electron gas surrounded by charge reservoirs. The same partitioning has been carried out in this case. The electron can enter and exit the 2D system at any random point (with some energy considerations), e.g. from the boundaries.}
	\label{universal-resist}
\end{figure*}
\begin{figure*}
	\begin{subfigure}{0.27\textwidth}\includegraphics[width=\textwidth]{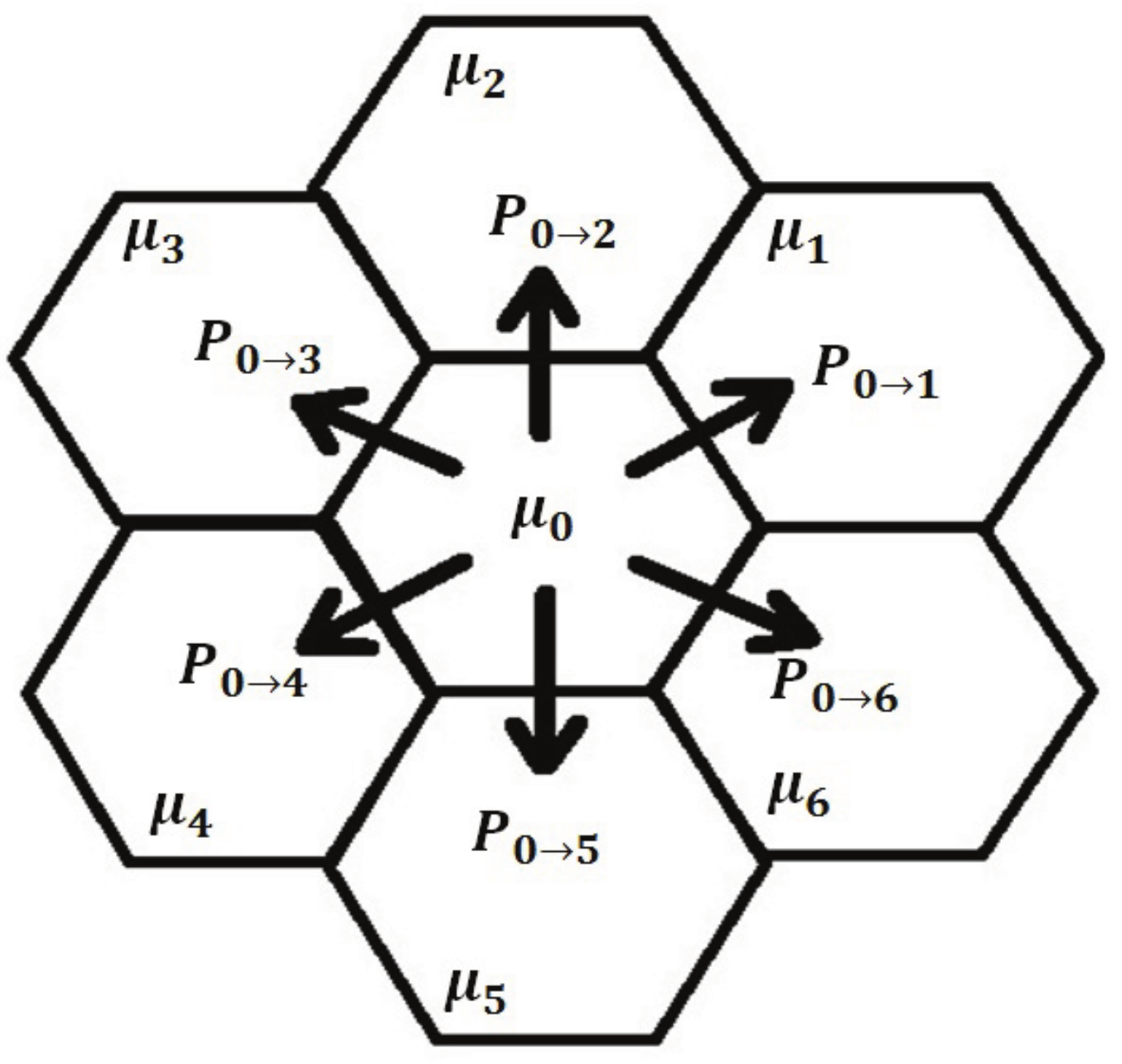}
		\caption{}
		\label{Transport}
	\end{subfigure}
	\begin{subfigure}{0.40\textwidth}\includegraphics[width=\textwidth]{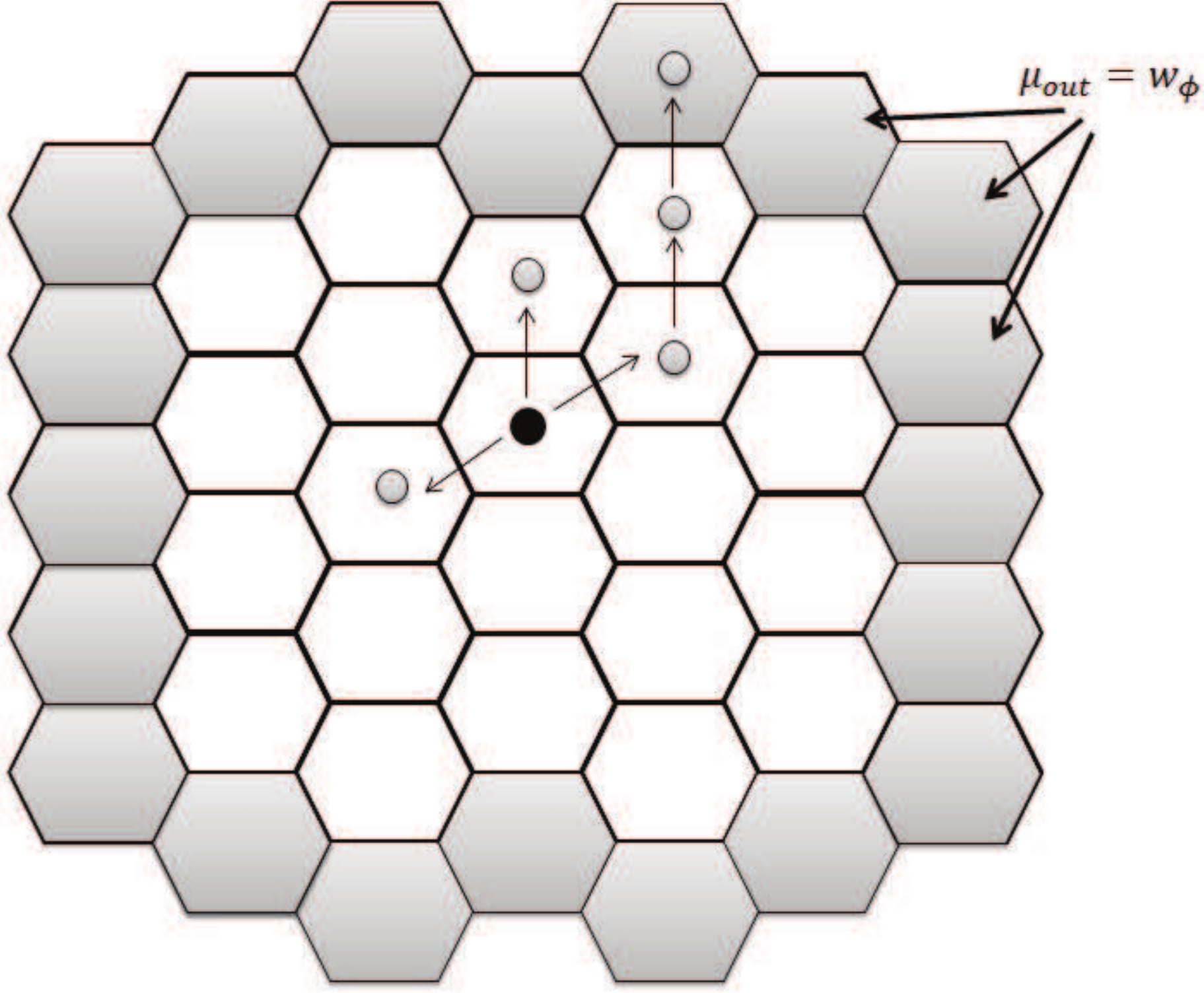}
		\caption{}
		\label{Movement}
	\end{subfigure}
	\caption{(a) The electron transition to the neighboring cells. (b) An schematic movement pattern of an electron in the virtual lattice. The gray cells are boundary sites (outer sites with $\mu=w_{\phi}$) from which the electrons can leave the system. The black circle is the site at which the electron has been injected and the gray circles represent the sites which have been unstable and relaxed through a chain of charge transfers}
	\label{MovementPattern}
\end{figure*}
Flicker noise (also called $1 / f$ noise) is a scale-free frequency fluctuations observed in physical systems, ranging from condensed matters to natural and economic systems. This noise was first observed in a vacuum tube in 1925 and then in many electrical devices and materials~\cite{johnson1925schottky}. The resistance fluctuations are present in resistors in the absence of driving current, as well as the dc and as currents. Scale invariance and independence of the underlying model (what we call universality) are two important features of $1/f$ noise in condensed matter systems. The synonym $1/f$ has originated from the fact that the power spectrum of the corresponding noise (reported firstly experimentally) in the form of $ S(f) \propto {f^{ - \alpha }}$, for which $0.8 < \alpha  < 1.4 $ ~\cite{weissman19881}. The effect of correlations in these noises has been a big question for a long time~\cite{dutta1981low}. There is a long-term debate about what mechanism is exactly behind this phenomena, e.g. whether it is deterministic or stochastic, whether the fluctuations are the carrier number or the mobility and surface or volume effect and the source of the (local) noise. Balandin et al.~\cite{liu2013origin} find a relationship between thickness and $1 / f$ noise in wire, concluding that if the thickness is more than seven atomic layers, it is a volume effect and otherwise it is a surface effect. There has been done a lot of research on these cases, which you can refer to ~\cite{hooge19761,hooge1981experimental,van1979flicker} for details. Based on experimental observations it was proposed that (Hooge's formula)~\cite{hooge19761,hooge1981experimental} 
\begin{equation}
\frac{{{S_R}(f)}}{{{R^2}}} = \frac{{{\alpha _H}}}{{{N_c}f}} 
\end{equation}
where $N_{c} $ is the number of carriers in a homogeneous sample, $ R $ resistors and ${{S_R}(f)} $ power spectral density of the fluctuations of the resistors at frequencies $f$ and ${\alpha _H} \approx 2 \times {10^3} $. The modified hooge's formula shows an acceptable agreement with the data related to semiconductors or Au-doped films with impurities, etc., provided that ${\alpha _H} $ is treated as an adjustable parameter~\cite{vandamme19831,hooge19691}. \\
Among many studies for explaining the flicker noise, one can mention the approaches based on temperature fluctuations~\cite{miller19811,voss1976flicker,weissman19881},
the Mcwhorter model (using the Lorentz spectra based on quantum tunneling in which the corner frequencies result from electronic traps )~\cite{bernamont1937fluctuations,mcwhorter1957semiconductor,van1979flicker}, universal conductance fluctuations in 1D and 2D due to the motion of interstitial impurities~\cite{feng1986sensitivity,weissman1987spectrum}, limiting case of broadened kinetics based on general arguments~\cite{weissman1981proceedings}, transport effect caused by the exchange of some conserved quantity (such as the energy or the carrier number) between the electronic system and an external reservoir~\cite{scofield1985resistance}, and the Dutta-Horn approach attributing the $1/f$ noise to the broad distribution of activation energies for the tunneling processes~\cite{dutta1979energy}. In the Dutta-Horn approach (which can simply be described by a two-state system with the energy difference $ \Delta E$ and the thermal activation energy ${E^ \pm } $, for which the tunneling rate is ${f_0}{e^{\frac{{ - {E^ \pm }}}{{kT}}}}$). The joint distribution of energy is considered to be $D({E^ \pm },\Delta E) $. Using the temperature-dependent occupancy variance and using the Lorentzian spectra with corner frequency ($f_c$) given by the sum of the two switching rates between the two states~\cite{weissman19881}, one finds that $ S(f,T) = \frac{{{{{\mathop{\rm sech}\nolimits} }^2}(\frac{{\Delta E}}{{2kT}})(\frac{{{f_c}}}{{f_c^2 + {f^2}}})}}{{2\pi }}$, using of which one comes up with the $1/f$ noise in this system. \\

The Dutta-Horn approach is based on the Helmholtz free energy differences, not only the energy, since the number of carriers fluctuates in each traped site, i.e. the entropy difference also plays a role. The relatively uniform $ {D_1}({E^ \pm })$ distribution function for $ {E^ \pm } \sim 20 + 2kT$ should be presented as follows in order to obtain a $1/f$ noise:
${D_1}({E^ \pm }) = \int\limits_{ - \infty }^{ + \infty } {D({E^ \pm },\Delta E){{{\mathop{\rm sech}\nolimits} }^2}} \left[ {\frac{{\Delta E}}{{2kT}}} \right]d\Delta E $. Using some reasonable conditions, Dutta-Horn proposed the following relation
\begin{equation}
\int\limits_f^\infty  {S(f')df' \propto \int\limits_0^{kT\ln (\frac{f}{{{f_0}}})} {{D_1}({E^ \pm })d{E^ \pm }} }
\end{equation}
where $S(f) = \frac{{{S_R}(f)}}{{{R^2}}}$.

\section{Our Approach:\\
Quantum Transition in Coherent Cells Formed by the phase relaxation Length}
\begin{figure*}
	\begin{subfigure}{0.45\textwidth}\includegraphics[width=\textwidth]{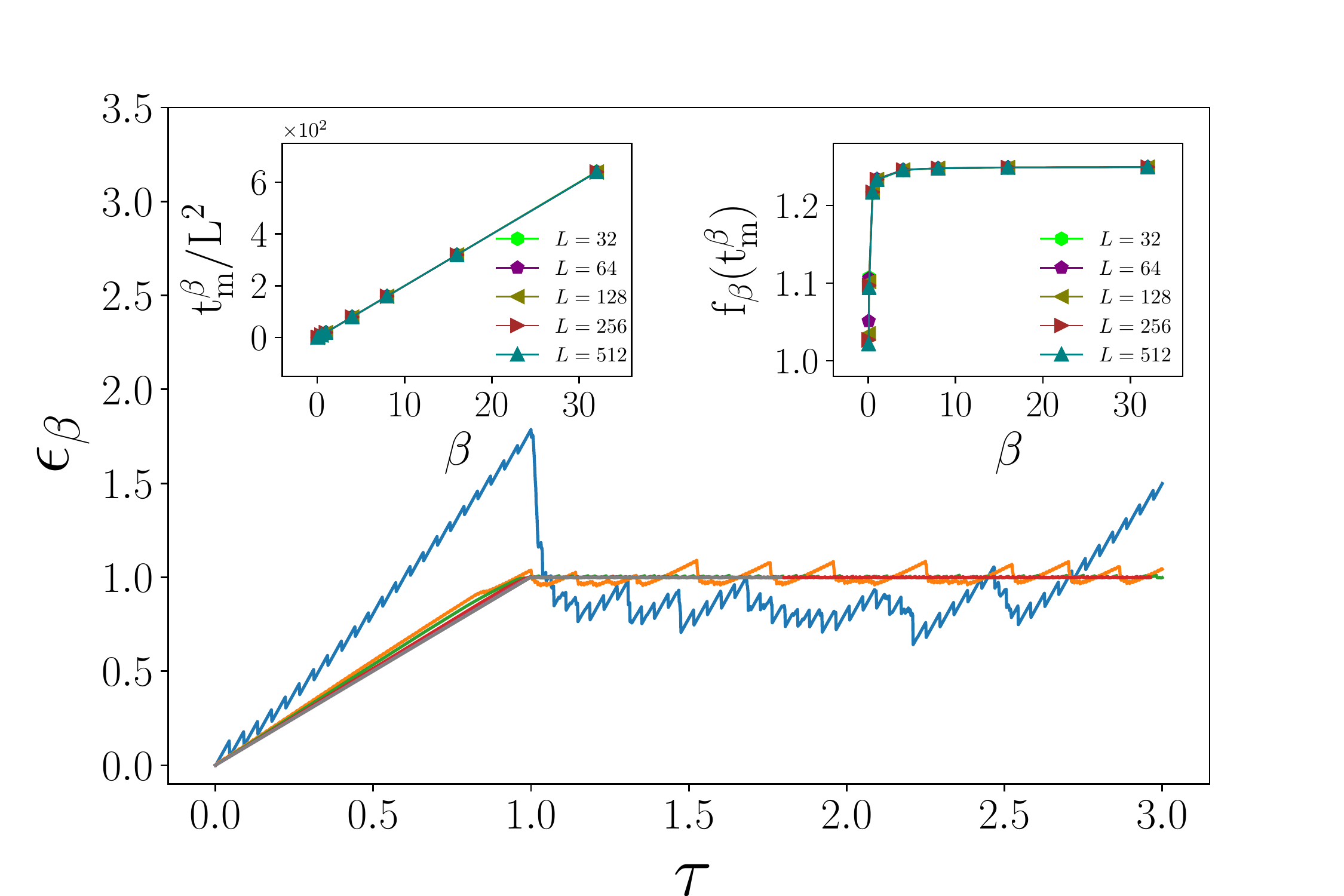}
		\caption{}
		\label{aaveL512betashif.pdf}
	\end{subfigure}
	\begin{subfigure}{0.45\textwidth}\includegraphics[width=\textwidth]{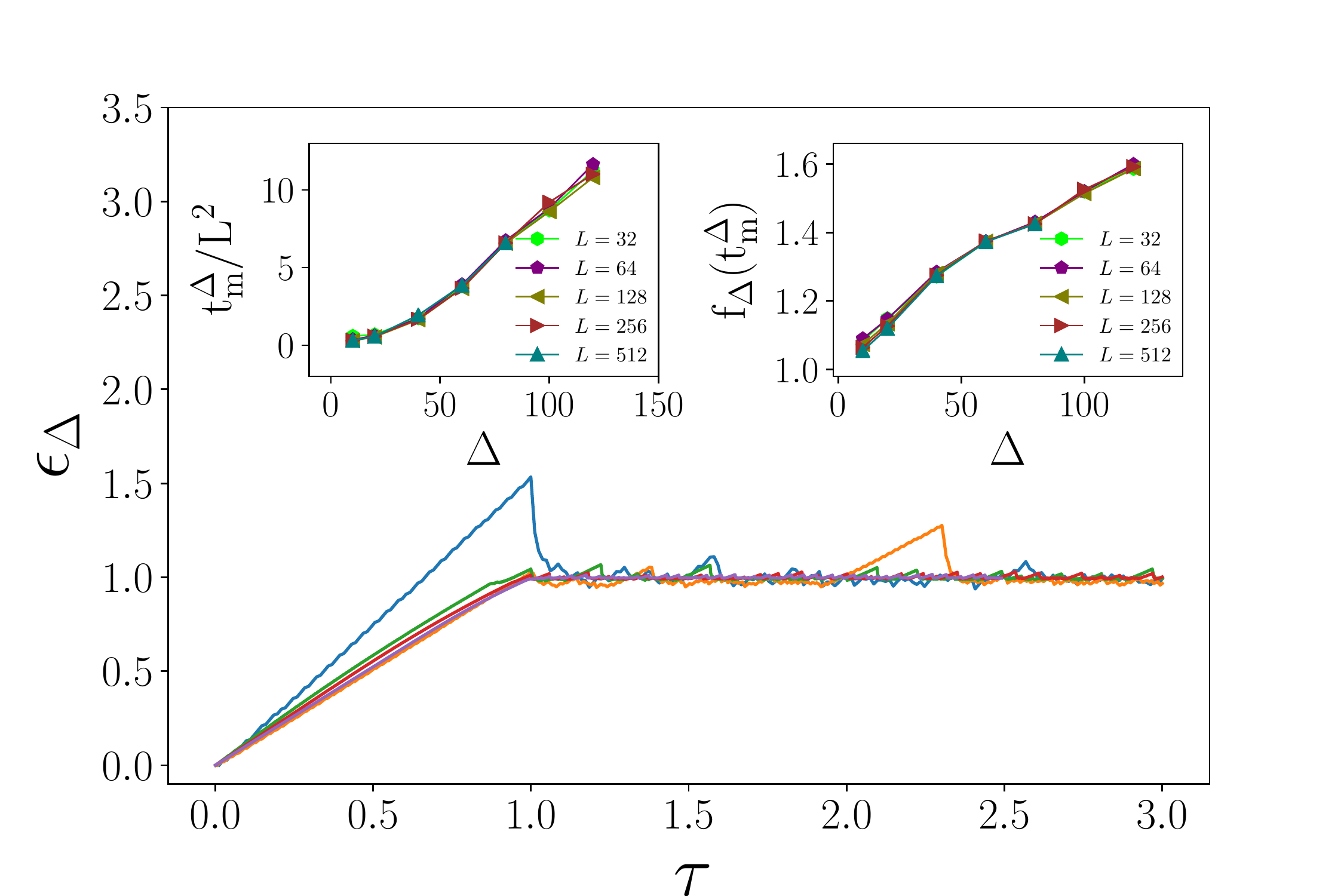}
		\caption{}
		\label{aaveL512deltashif.pdf}
	\end{subfigure}
	\caption{(Color Online) (a) ${\epsilon _\beta }$ in term of time $t$ for various $\beta$ for $L=512$. Left inset: $\frac{{t _m^\beta }}{{{L^2}}}$ in term of $\beta$ for various $L$. Right inset: ${f_\beta }(t _m^\beta )$ in term of $\beta$ for various $L$. (b) ${\epsilon _\Delta }$ in term of time $t$ for various $\beta$ for $L=512$. Left inset: $ \frac{{t _m^\Delta }}{{{L^2}}} $in term of $\Delta$ for various $L$. Right inset: ${f_\Delta }(t _m^\Delta )$ in term of $\Delta$ for various $L$. }
	\label{fig:stationarity}
\end{figure*}
\begin{figure*}
	\begin{subfigure}{0.45\textwidth}\includegraphics[width=\textwidth]{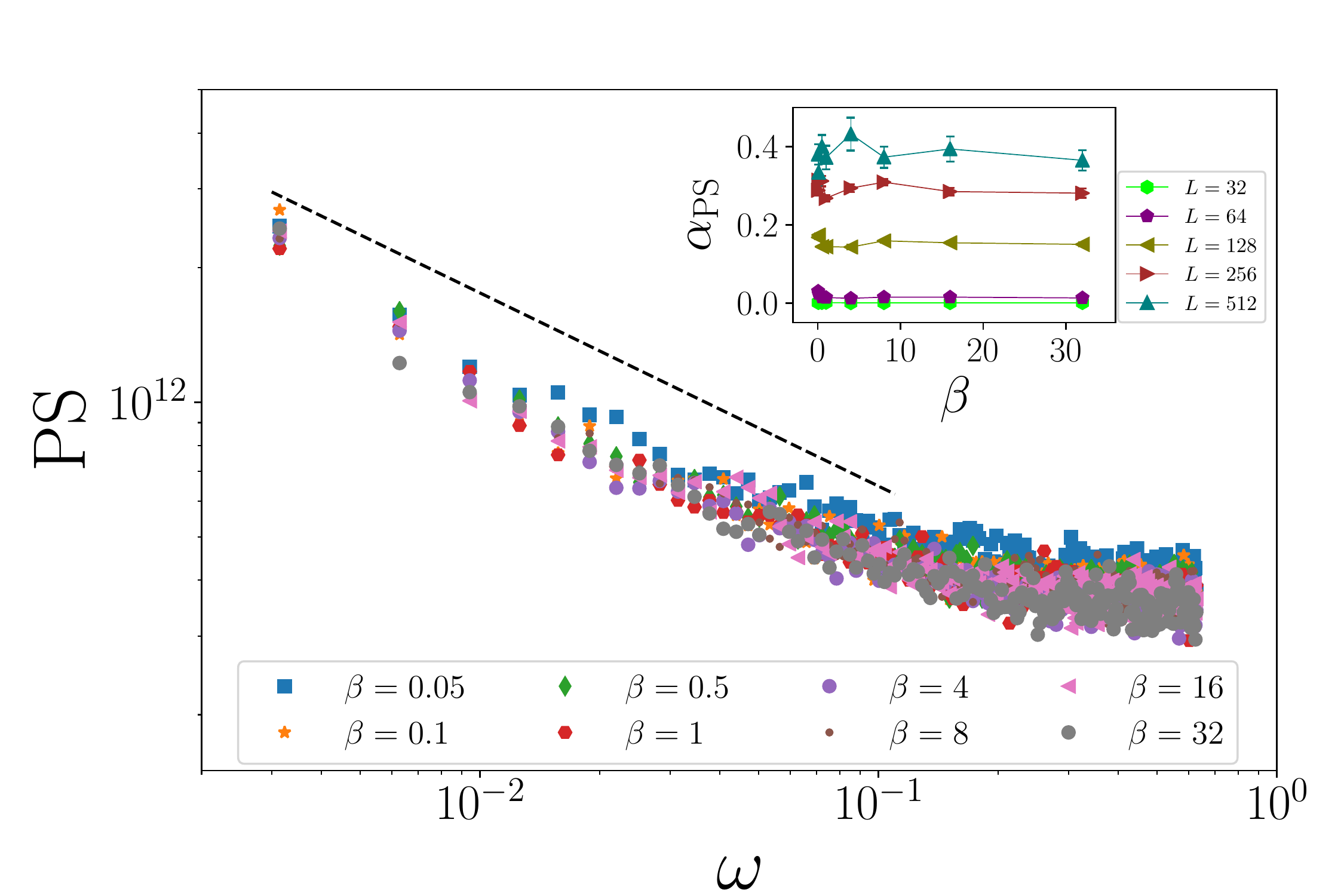}
		\caption{}
		\label{PS(s)512bs.pdf}
	\end{subfigure}
	\begin{subfigure}{0.45\textwidth}\includegraphics[width=\textwidth]{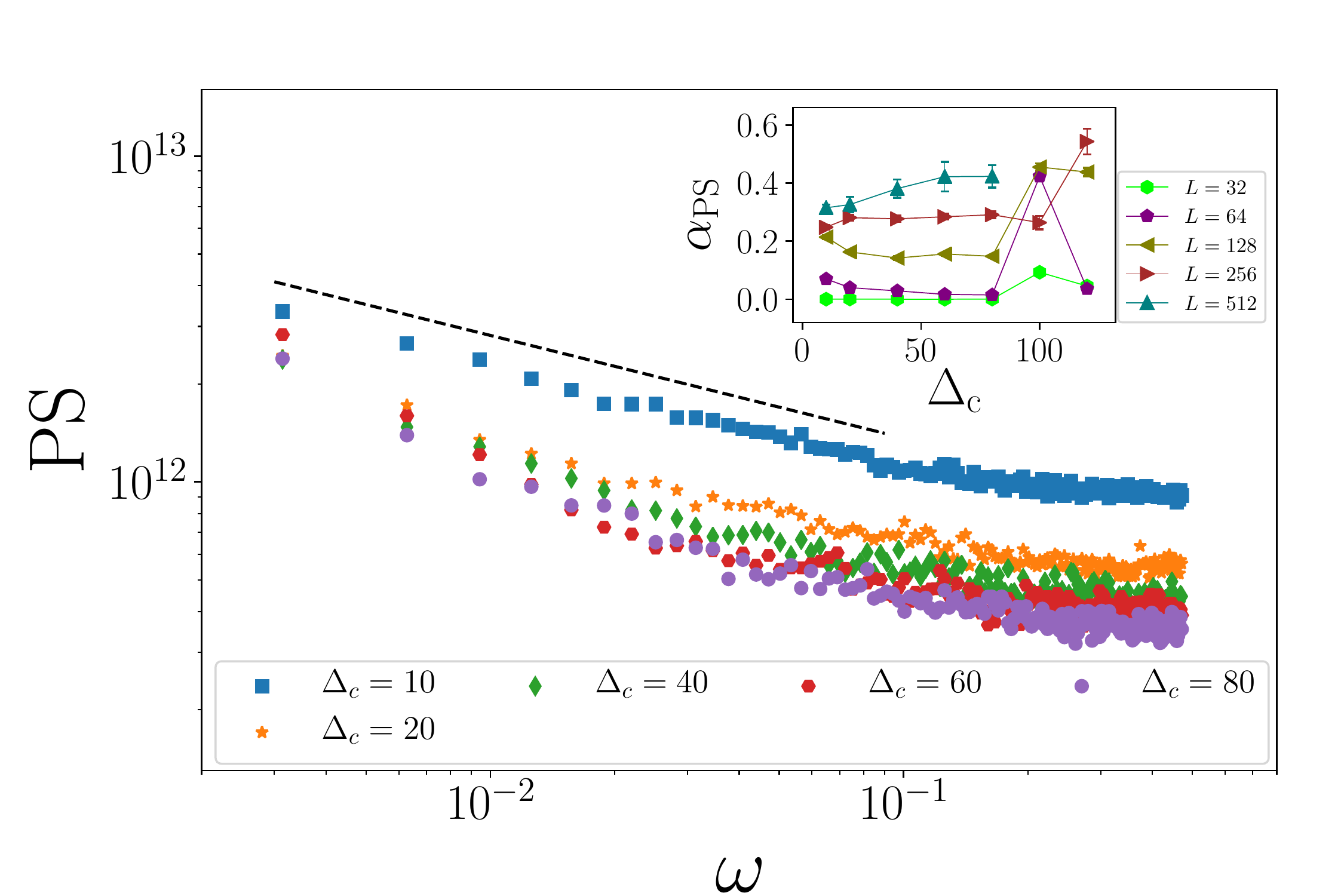}
		\caption{}
		\label{PS(s)512ds.pdf}
	\end{subfigure}
	\caption{(Color Online) (a) Power spectrum in term of $\omega $ for various $\beta$ for $L=512$. Inset: $\alpha_{PS}$ in term of $\beta$ for various $L$. (b) Power spectrum in term of $\omega $ for various $\Delta$ for $L=512$. Inset:  $\alpha_{PS}$ in term of $\Delta$ for various $L$.}
	\label{fig:aave512}
\end{figure*}

\begin{figure}
	\begin{subfigure}{0.45\textwidth}\includegraphics[width=\textwidth]{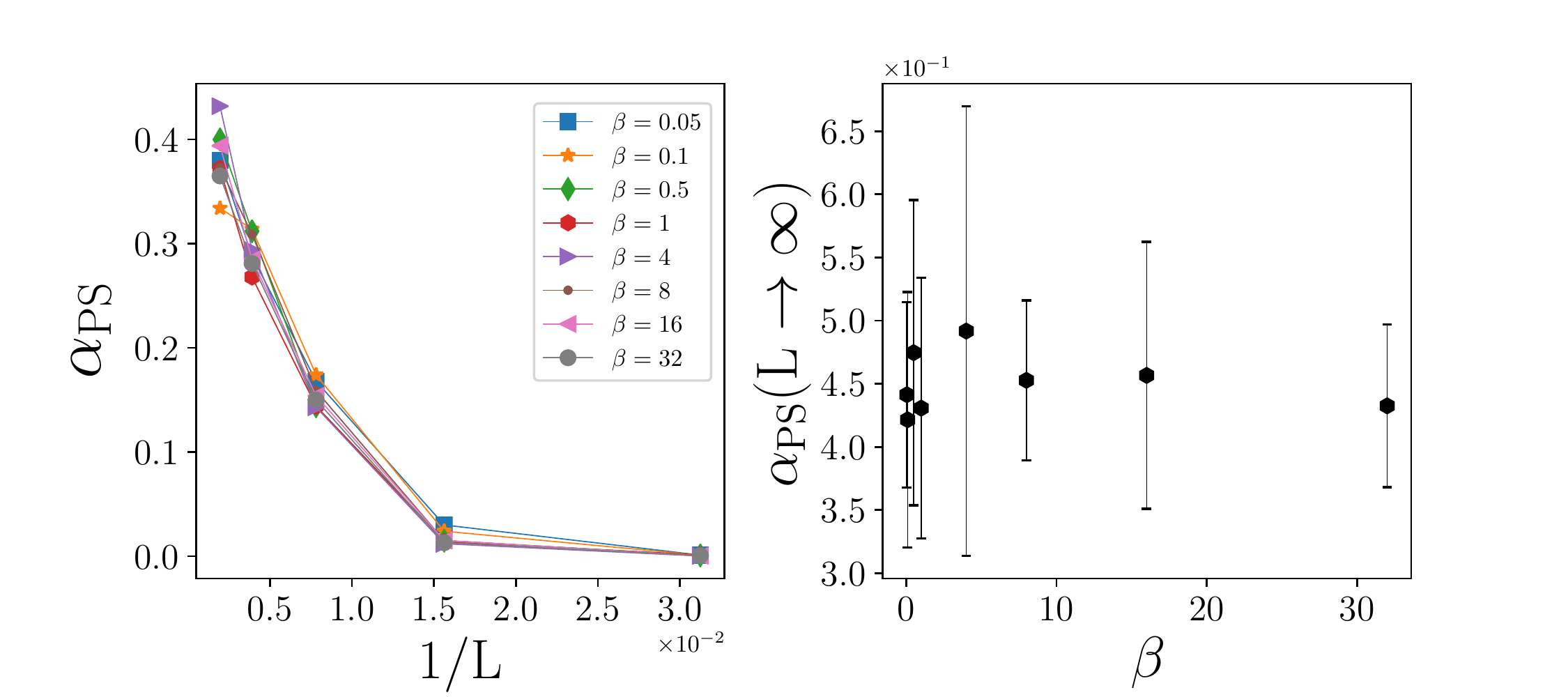}

		\caption{}
		\label{slopePSbs.pdf}	
	\end{subfigure}
	\begin{subfigure}{0.45\textwidth}\includegraphics[width=\textwidth]{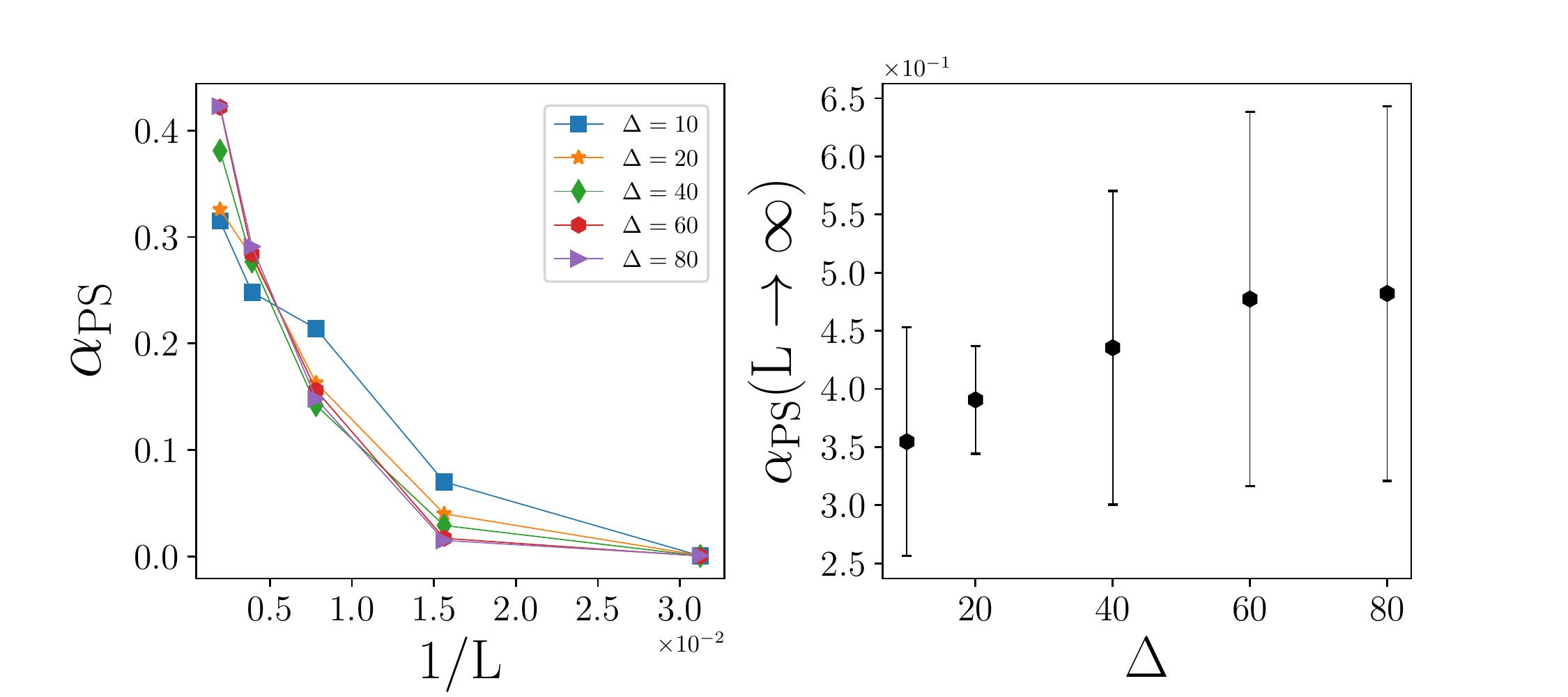}

		\caption{}
		\label{slopePSds.pdf}	
	\end{subfigure}
	\caption{(a) ${\alpha _{PS}}$ in term of $\frac{1}{L}$ for $L=512$ that fit with $a{x^2} + bx + c$ and report $c$ ($ \alpha_{PS} (L \to \infty ) $) in term of $\beta$. (b) ${\alpha _{PS}}$ in term of $\frac{1}{L}$ for $L=512$ that fit with $a{x^2} + bx + c$ and report $c$ ($ \alpha_{PS} (L \to \infty ) $) in term of $\Delta$.}
	\label{fig:slopeb512}
\end{figure}
In this section we briefly describe the method that was developed in~\cite{najafi2018percolation,najafi2019electronic}. In this dynamical model, percolation of electrons with avalanche dynamics was proposed as the source for MIT of a two-dimensional electron gas at zero magnetic field. The model successfully explained major features of 2D MIT in semiconductor-Oxide-metal junction, mainly a transition line between a percolating and a non-percolating phase was discovered. The percolating phase was shown to be metallic phase in which the mobility (proportional to the spanning avalanche probability $\sigma$) is decreasing function of temperature ($\frac{\text{d}}{\text{d}T}\sigma<0$), while for the non-percolating phase the inverse (insulating phase) was observed. \\

In this model, we consider a 2DEG in contact with an electronic reservoir. As a well-known fact, in the normal phase (in the sense of the Anderson localization), where the electrons undergo normal diffusion, the phase relaxation length behaves like $ {\zeta_\phi } = \sqrt {D{\tau _\phi }} $, where $D$ is diffusion coefficient and $ {\tau _\phi } $ is the phase relaxation time above which the electron losses its \textit{quantum memory}~\cite{altshuler1985electron}. It depends on temperature, the electron-electron interaction (via the mean free length $l$) and the disorder strength. Accordingly, two important types (scales) of dynamics for electrons is formed with respect to the phase relaxation length: $l \ll r \ll \zeta_{\phi}$ and $r\gg \zeta_{\phi}$, where $r$ is the length scale of the electron dynamics which in the diffusive regime behaves like $r\sim\sqrt{Dt}$ with time $t$. To be more precise let us show the state of an electron in the initial time by $\psi=\sqrt{\rho}e^{-\alpha(0)}$ where $\rho$ shows the density, and $\alpha(t)$ is its phase which change over time. Then for $t\ll\tau _\phi$ the phase of $\psi$ is almost unchanged (the first regime, where the electron sustains its quantum phase, while for $t\gg\tau_\phi$ or equivalently $r\gg \zeta_{\phi}$ the quantum phase of the electrons becomes nearly random, so that a semi-classical transport is applicable, i.e. the quantum fluctuations do not play a vital role. This approach has been proved to be very useful for the interpretation of finite-size power-law conductivity of 2DEG~\cite{backes2015observation}, the self-averaging~\cite{bruus2004many} and percolation prescription of 2DEG~\cite{meir1999percolation}, each of which considers the linear size  $\Delta L\sim\zeta_{\phi}$ as an important spatial scale~\cite{najafi2019electronic}. \\

The main building block of our model is to divide 2D system to many \textit{cells} (called coherent cells) each of which has linear size $\zeta_{\phi}$. These regions are expected to form around the impurities/disorders. The electrons gas inside the cells are treated purely quantum mechanically as the coherent electron gas. The transport of electrons between the coherent cells (or simply cells hereafter) are based on electronic tunneling according to the difference between the free energies. For this last part we use the Metropolis Monte-Carlo method which is, in esprit (semi) classical. We consider the cells to be hexagonal to be as symmetric as possible, and at the same time simple to analyze. In this model, electrons are dispersed through the system according to the (temperature-dependent) energy content as well as the chemical potentials of the cells. Upon adding an electron to the system, the density of electron in all sites of the system is updated according the laws to be explained in the rest of this section. The system under study is schematically shown in Fig.~\ref{fig:sample1} and~\ref{fig:Honeycombb}, the latter visualizing the experimental setup where a 2DEG is formed between the layers of metal-oxide-semiconductor multi-layer. The electrons are permitted to enter from/leave the system to the reservoirs. The colored region shows an \textit{electronic avalanches}, see the followings for the details.

\subsection{The Free Energy and the Transition Rules} 
\begin{figure*}
	\begin{subfigure}{0.48\textwidth}\includegraphics[width=\textwidth]{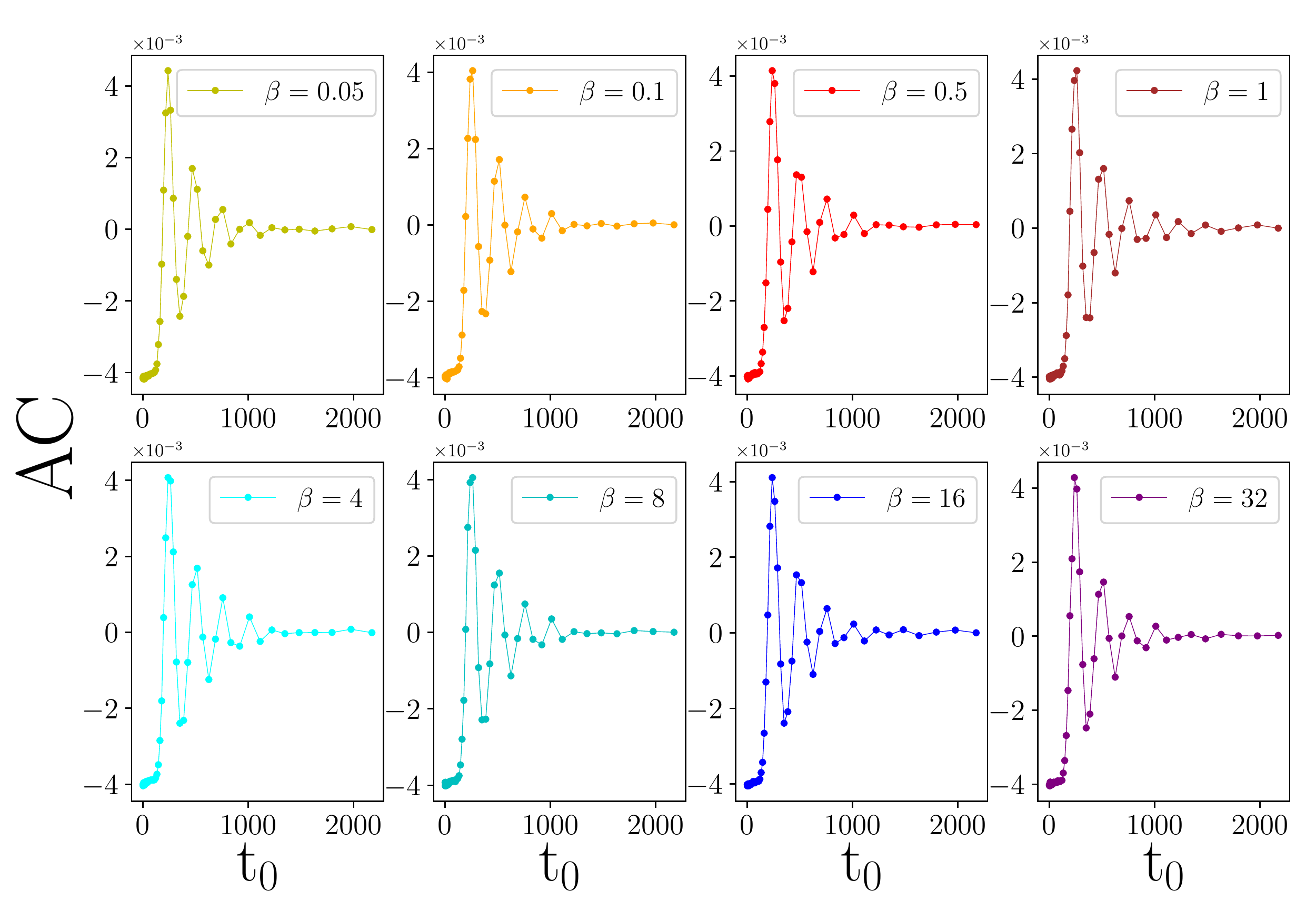}
		\caption{}
		\label{fig:Ab32}
	\end{subfigure}
	\begin{subfigure}{0.48\textwidth}\includegraphics[width=\textwidth]{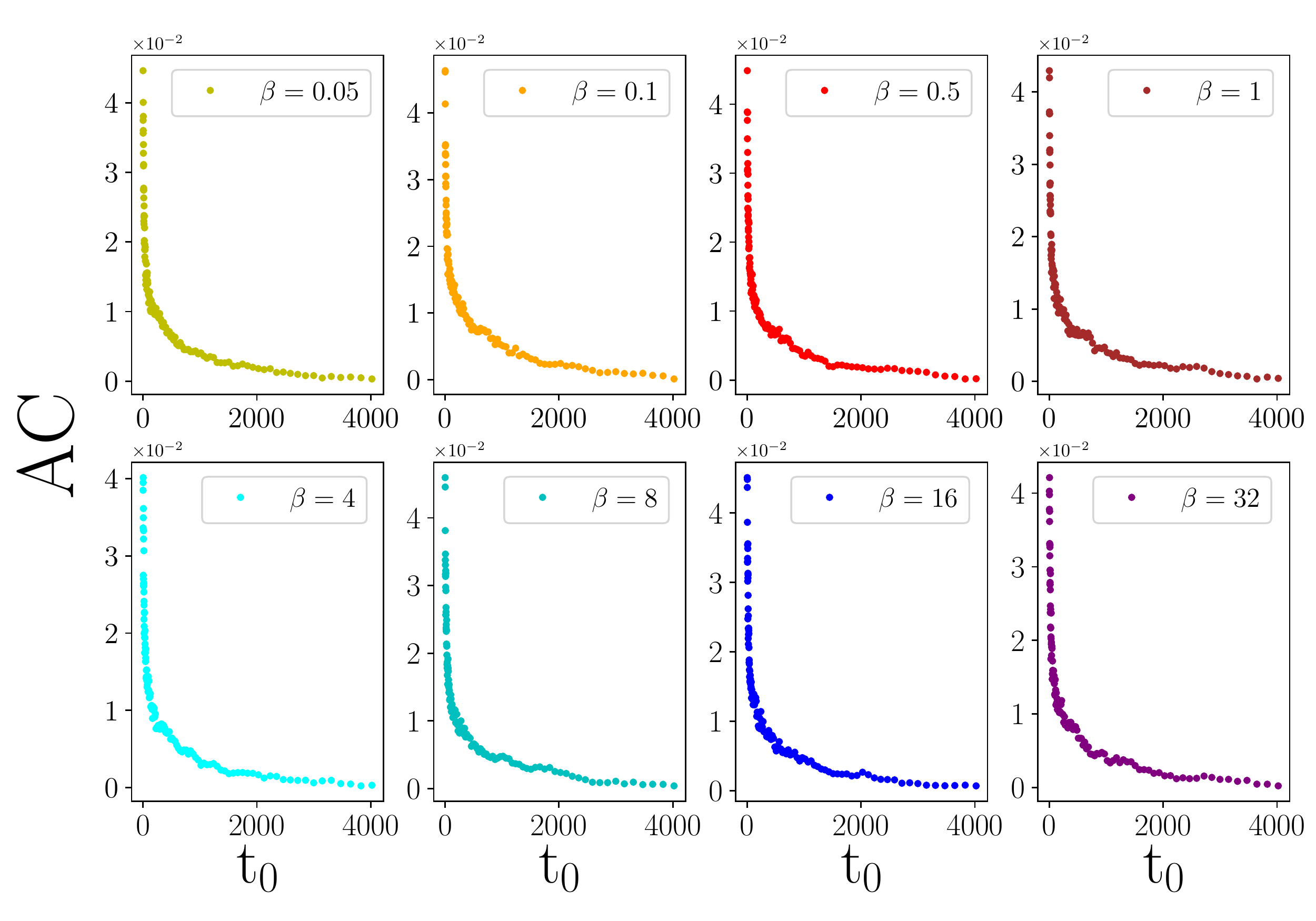}
		\caption{}
		\label{fig:Ab512}
	\end{subfigure}
	\begin{subfigure}{0.55\textwidth}\includegraphics[width=\textwidth]{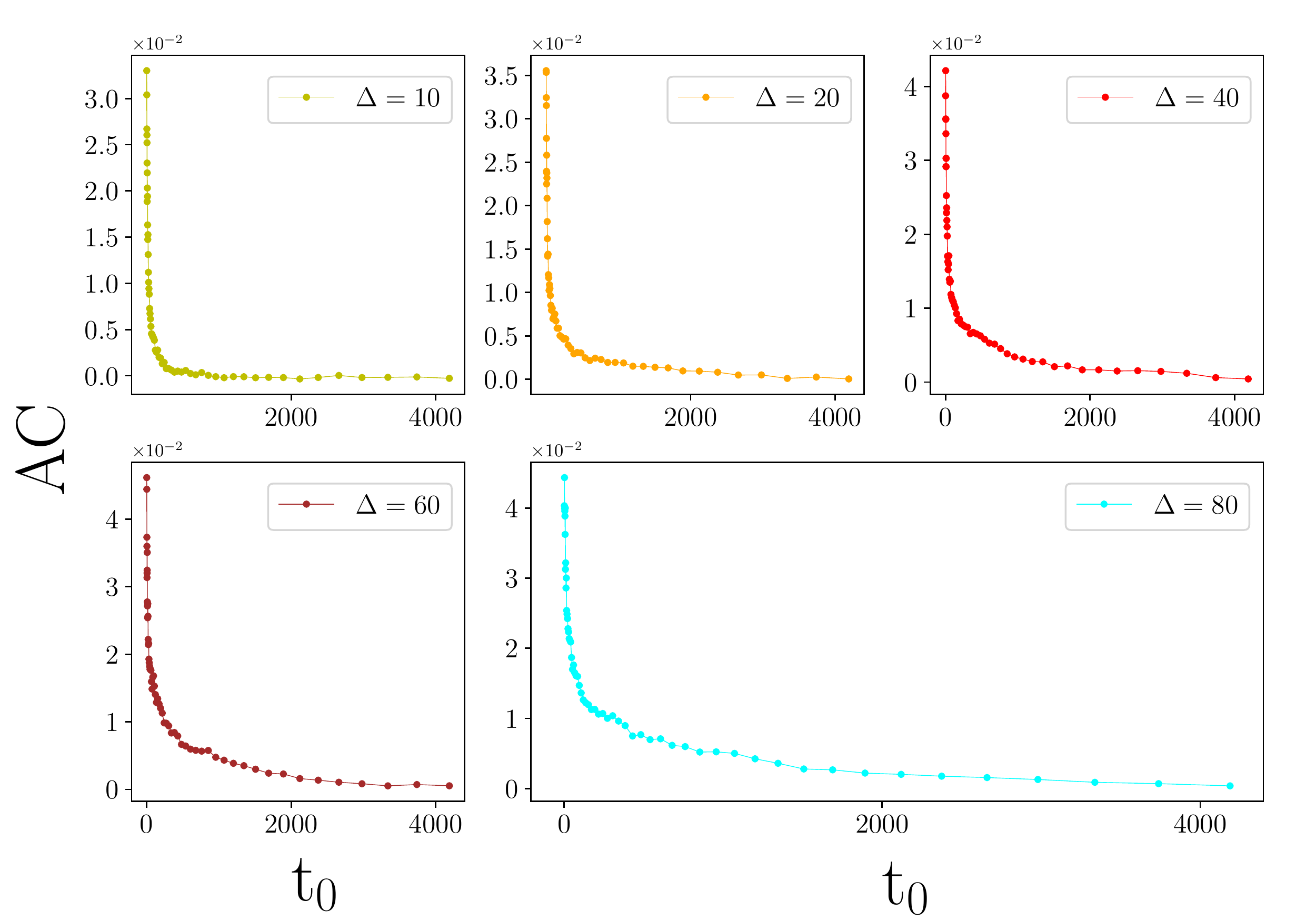}
		\caption{}
		\label{fig:Ad256}
	\end{subfigure}
	\caption{(a) Autocorrelation in term of $t_{0}$ for various $\beta$ for $L=32$. (b) Autocorrelation in term of $t_{0}$ for various $\beta$ for $L=512$. (c) Autocorrelation in term of $t_{0}$ for various $\Delta$ for $L=512$.  }
	\label{fig:Auto}
\end{figure*}
In~\cite{najafi2019electronic} it was shown that the probability of adding an electron to $i$th cell is proportional to
\begin{equation}
\begin{split}
p_{\text{adding particle}}&\propto\exp\beta\left[ \mu_0-(A_{\tilde{N}_i+1}(V,T)-A_{\tilde{N}_i}(V,T))\right]\\
& = \exp\beta\left[ \mu_0-\mu_{\tilde{N}_i}(V,T)\right].
\end{split}
\label{relativProb}
\end{equation}
where $A_{\tilde{N}_i}(V,T)$ is the Helmholtz free energy of $i$th cell with $\tilde{N}_i$ electrons and $\mu_{\tilde{N}_i}(V,T)\equiv \frac{\partial A}{\partial \tilde{N}_i}$, $\beta=\frac{1}{k_BT}$ is the inverse of temperature, and $\mu_0$ is the average chemical potential of the total system. The equation of the equilibrium state is given by $ \mu_0=\mu_{\tilde{N}_i}\equiv\mu_i$. Also the probability of transferring an electron from the $i$th cell to its neighbor $j$ is proportional to $\exp\left[-\beta(\mu_j-\mu_i)\right]$~\cite{najafi2019electronic}. For a given configuration of chemical potentials, it is easy to find the transport rules. We identify the system by a configuration of the chemical potentials $\left\lbrace\mu_i\right\rbrace_{i=1}^N$ throughout the system. A cell $i$ is called unstable, if $\mu_i>\mu_0$ after which the electrons in it are candidates to move to the neighboring cells. The local rule for this transition, which is schematically shown in Fig~\ref{Transport} is simply by looking at the difference between the chemical potentials: for this figure the difference between $\mu(0)$ (as the donor) and $\mu(j)$, $j=1,2,...,6$. Without loose of generality, let us suppose that $\mu(i)<\mu(j)$ for $i<j$, then the transition takes place first towards the site with lowest $\mu$ content, i.e. $\mu(1)$. The probability of such a transition in the classical Metropolis scheme is given by
\begin{equation}
	P_{0\rightarrow j}\sim \Theta(\mu(0)-\mu_0) \times\text{Max}\left\lbrace 1,e^{-\beta\left( \mu(j)-\mu(0)\right)} \right\rbrace.
	\label{Eq:transport_prob}
\end{equation}
where $\Theta(x)$ is the step function, guaranteeing that the donor site is unstable.\\

The next essential step is to find the chemical potentials according to which the dynamics is defined. To this end, one should find the energy functional and the free energy of the system. The electron gas energy and chemical potential within each cell (inside which the density is \textit{supposed to be uniform}) was calculated using the Thomas-Fermi-Dirac (TFD) approach in~\cite{najafi2019electronic}. The uniform average energy of the $i$th cell is given by $\left\langle E_i \right\rangle=K(T,\tilde{N}_i)+V_{ee}(T,\tilde{N}_i)+E_{\text{imp}}(T,\tilde{N}_i)$ in which the terms are finite temperature averages of the kinetic, the electron-electron interaction and the impurity energies respectively and $\tilde{N}_i$ is the number of electrons in the cell. The total energy of a cell is shown to be~\cite{najafi2019electronic}:
\begin{equation}
E_T=\sum_{i=1}^{L}\left[-\alpha T^2\text{Li}_2\left(1-e^{N_i/T}\right) +\beta_0 N_i^2-\gamma_i N_i\right]
\end{equation}
in which $m$ is the electron mass, $\epsilon_0$ is the vacuum dielectric constant, $Z_i$ is the random disorder in the $i$th cell, $\alpha=2m\left( \frac{\pi k_B \zeta_{\phi}(T)}{\hbar}\right)^2$, $\beta_0=\frac{1}{8\sqrt{2}\epsilon_0 \zeta_{\phi}(T)}\left( \frac{e\alpha}{k_B}\right)^2$, $\gamma_i=\text{sinh}^{-1}(1)\frac{\alpha e^2}{\pi\epsilon_0k_B \zeta_{\phi}(T)}Z_i$, $N_i=\frac{k_B}{\alpha}\tilde{N}_i$ and $L$ is the total number of cells. In this relation $Li_{2}$ is the polylogarithm function of order $2$, and $Z$ is picked from the distribution $P(Z) = \frac{1}{\Delta }\Theta (\frac{\Delta }{2} + (Z - {Z_0})\Theta (\frac{\Delta }{2} - (Z - {Z_0}))) $ where $\Delta$ is the disorder strength, and $Z_0$ is the average number of impurities. We have $ \left\langle Z \right\rangle  = {Z_0}$ and $\left\langle Z_iZ_j \right\rangle  = \delta _{ij} $, where $\delta$ is the Kornecker delta, and $\left\langle \right\rangle$ shows the ensemble average. The chemical potential of a cell is then obtained using the relation $A_{\tilde{N}}(V,T)-T\left(\frac{\partial A}{\partial T}\right)_{\tilde{N},V}=\left\langle E\right\rangle$. It is shown in~\cite{najafi2019electronic} that
\begin{equation}
\mu_i =k_BT\ln\left(e^{h_i}-1\right)+UT^{\frac{1}{2}}h_i-IZ_iT^{\frac{1}{2}}
\end{equation}
in which $U =\frac{2k_Bm\sqrt{2Da}e^2\pi^2}{8\epsilon_0\hbar^2}$, $I=\text{sinh}^{-1}(1)\frac{e^2}{\pi\epsilon_0\sqrt{Da}}$, $h_i=\frac{N_i}{T}$ and $i$ stands for the $i$th cell. The well-known fact $ \zeta_{\phi}(T)=aT^{-1/2}$ was used in this relation, which defines $a$. The effect of randomness of $Z_i$'s (that are supposed to be random noise with an uniform probability measure), which captures the on-site (diagonal) disorder was investigated in~\cite{najafi2019electronic}. Therefore, one finds the difference between the chemical potentials is given by 
\begin{equation}
\mu_2-\mu_1=k_BT\ln\left(\frac{e^{h_2}-1}{e^{h_1}-1}\right) +UT^{\frac{1}{2}}(h_2-h_1)-IT^{\frac{1}{2}} (Z_2-Z_1).
\end{equation}
which is used to determine the transfer probability. \\
\begin{figure*}
	\begin{subfigure}{0.45\textwidth}\includegraphics[width=\textwidth]{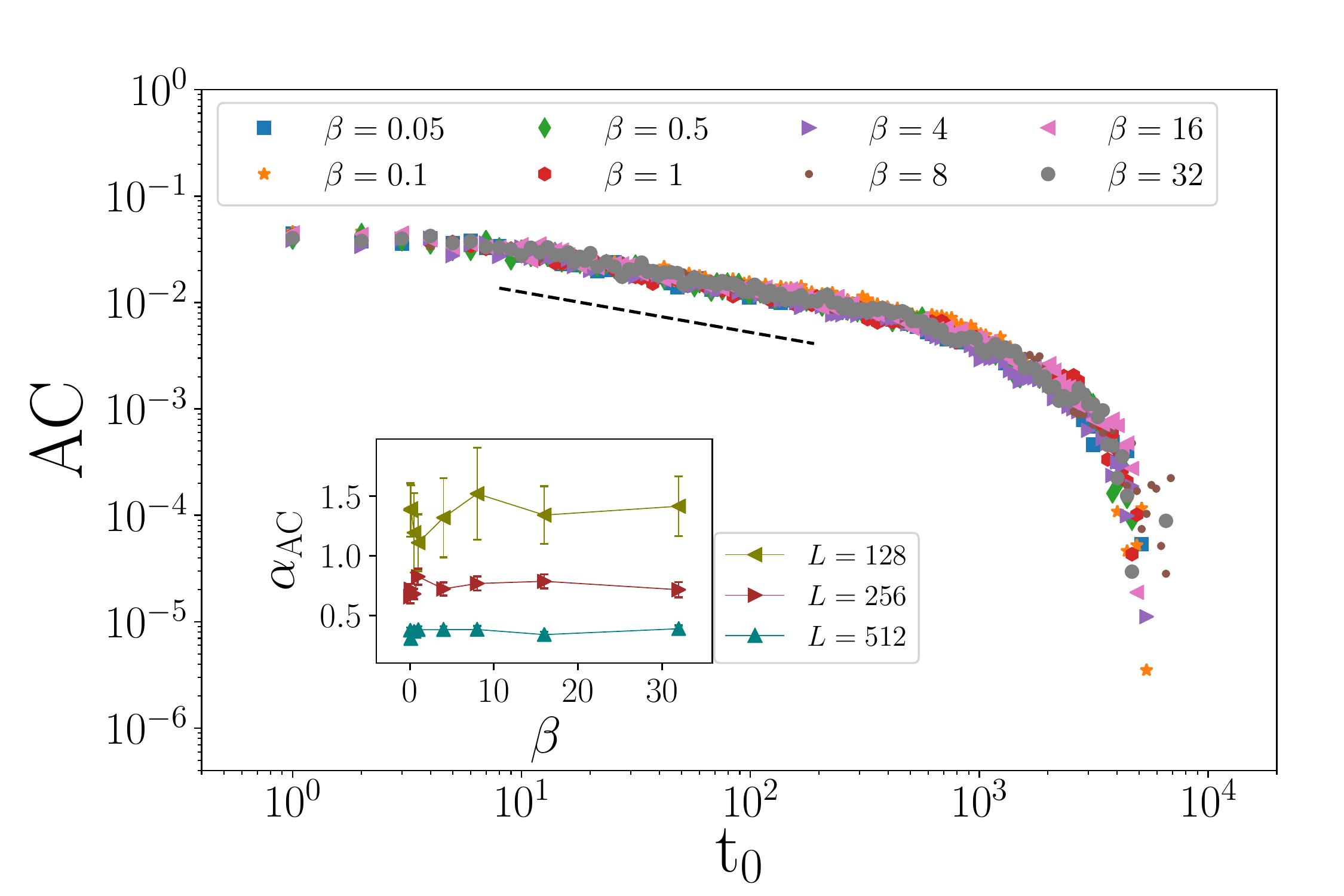}
		\caption{}
		\label{fig:A25121}
	\end{subfigure}
	\begin{subfigure}{0.45\textwidth}\includegraphics[width=\textwidth]{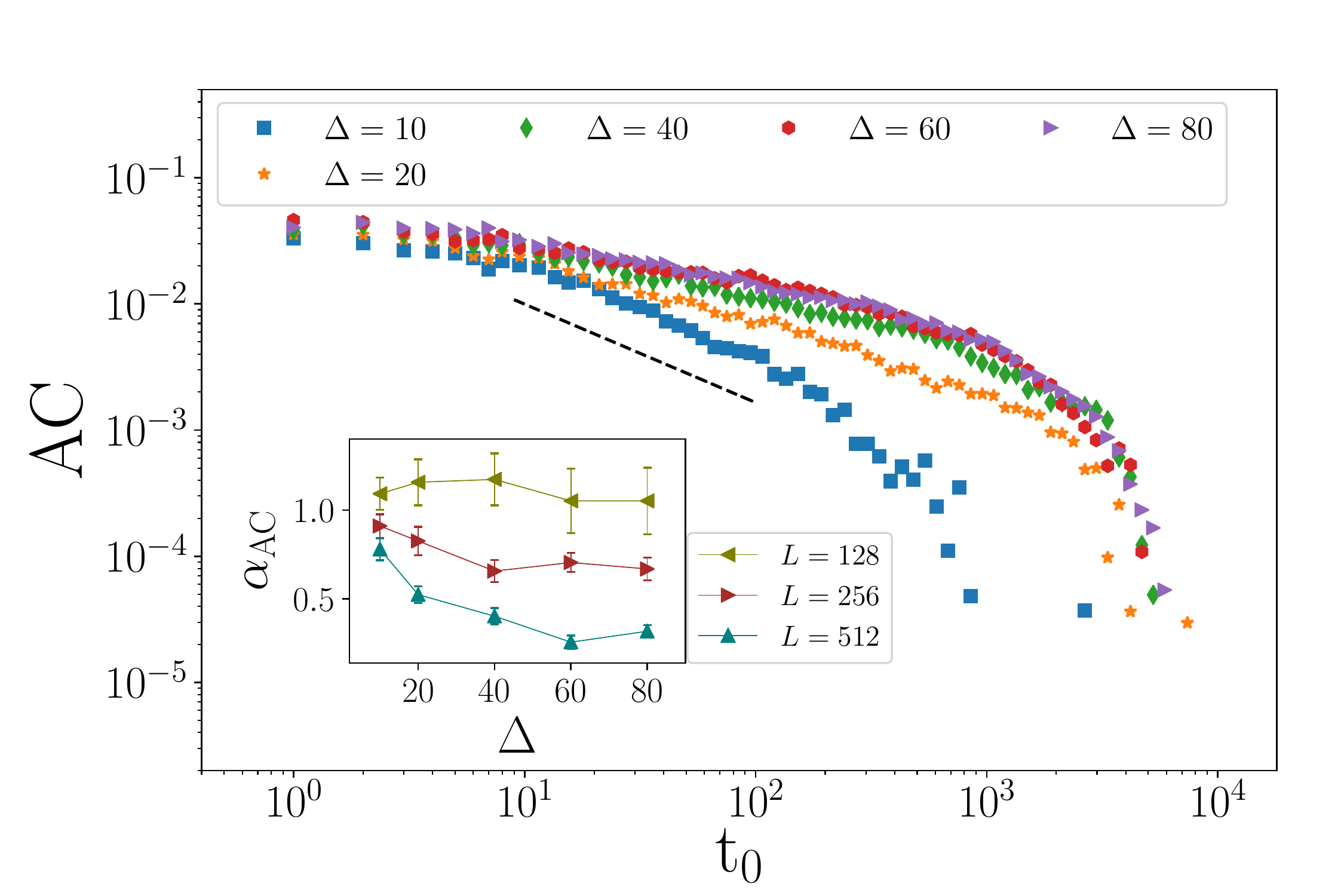}
		\caption{}
		\label{fig:A25122}
	\end{subfigure}
	\caption{(a) Autocorrelation in term of $t_{0}$ for various $\beta$ for $L=512$. Inset:$ {\alpha _{AC}} $ in term of $\beta$ for various $L$.\\ 
		(b) Autocorrelation in term $t_{0}$ for various $\Delta$ for $L=512$. Inset:$ {\alpha _{AC}} $ in term $\Delta$ for various $L$. }
	\label{fig:A2512}
\end{figure*}

Before explaining the Monte Carlo method for simulating the system, let us define some processes: \\
** A \textit{local equilibration} (or \textit{electronic toppling}) of a typical site $i$, is defined as the process in which the electrons of this site transfer to the neighbors (with a priority of sites with lower chemical potentials) according to Eq.~\ref{Eq:transport_prob} without the $\Theta$ function in front. In each local relaxation (electronic toppling) seven transitions are checked: six neighboring cells, and one transition to an external electronic source.  The order of neighboring cells for electronic transit is according to the order of their chemical potential as explained above.\\
** The \textit{equilibration process} is defined in which we apply \textit{locally equilibration} for some random sites $n$ times (where $n$ should be large enough to have better equilibrated system). \\
** The \textit{electron injection} is the process in which an electron is injected to a site. The \textit{relaxation process} defined as the process in which \textit{all} the unstable sites are locally equilibrated, after which no site is unstable. \\
** An \textit{electronic avalanche} is defined as the process which starts by a single injection (leading to an unstable site) up to the time where no site is unstable, i.e. the process in which the system starts from and ends in a stable configuration. \\
** The \textit{avalanche size} is the number of electronic topplings in an avalanche. \\
** It is notable that for some extreme conditions and circumstances there are very rare states, called \textit{extended} meaning that the system cannot become stable. In these cases we let the unstable configurations be in final states. Therefore, the above-mentioned definition of avalanches is not applicable.\\
An \textit{extended avalanche} is an avalanche whose size is larger than $50L^2$. \\ 

Now we explain the systematic way of simulating the system based on the Monte Carlo method. We consider $L\times L$ system, the state of which is identified by the configuration of two fields: $\left\lbrace h_i,Z_i\right\rbrace_{i=1}^{N=L^2}$, where $Z_i$ is random charged disorder which is taken from the distribution  $P(Z) = \frac{1}{\Delta }\Theta (\frac{\Delta }{2} + (Z - {Z_0})\Theta (\frac{\Delta }{2} - (Z - {Z_0}))) $ and $h$ is related to the carrier density (see above). Initially we set $h_i$ and $Z_i$ randomly in such a way that no site is unstable, i.e. $\mu_i<\mu_0$ for all $i\in \left\lbrace 1,2,...,L^2\right\rbrace $. Then an overall equilibration process is performed during which $n=10L^2$ sites are randomly selected and locally equilibrated. Then electron injection process begins in such a way that every time a random site is selected for injection, and if the resulting chemical potential of the selected site becomes greater than $ \mu_{0}$, the site is locally equilibrated. As a results an electronic avalanche is triggered with lat's say mass $m$. When the avalanche is over, an overall equilibration process takes place during which $m$ local equilibraions occure. For every $2000$ triggered avalanches, we impose a large scale equilibration in which $10L^2$ local equilibrations are done. In case of extended avalanche we stop the process, and impose a large scale equilibration.\\

Then we apply an equilibration to equilibrate the system for the first time. Then we select a random site for injecting an electron, after which we apply the relaxation process to have a chain of topplings over the system and reach a stable configuration. Note that the electrons may leave the system in two ways tuned by two work functions: from the bulk to the reservoir tuned by ${w_ \bot }$, and from the boundaries of the system tuned by ${w_\parallel }$. After each relaxation of the system, we apply an equilibration process to reach an equlibrated configuration. Note that generally the time scale of electronic injections is much larger than the time scale of equilibration. After this process, another site is selected for the electronic injection, and the process continues.

\section{simulation}
\begin{figure*}
	\begin{subfigure}{0.40\textwidth}\includegraphics[width=\textwidth]{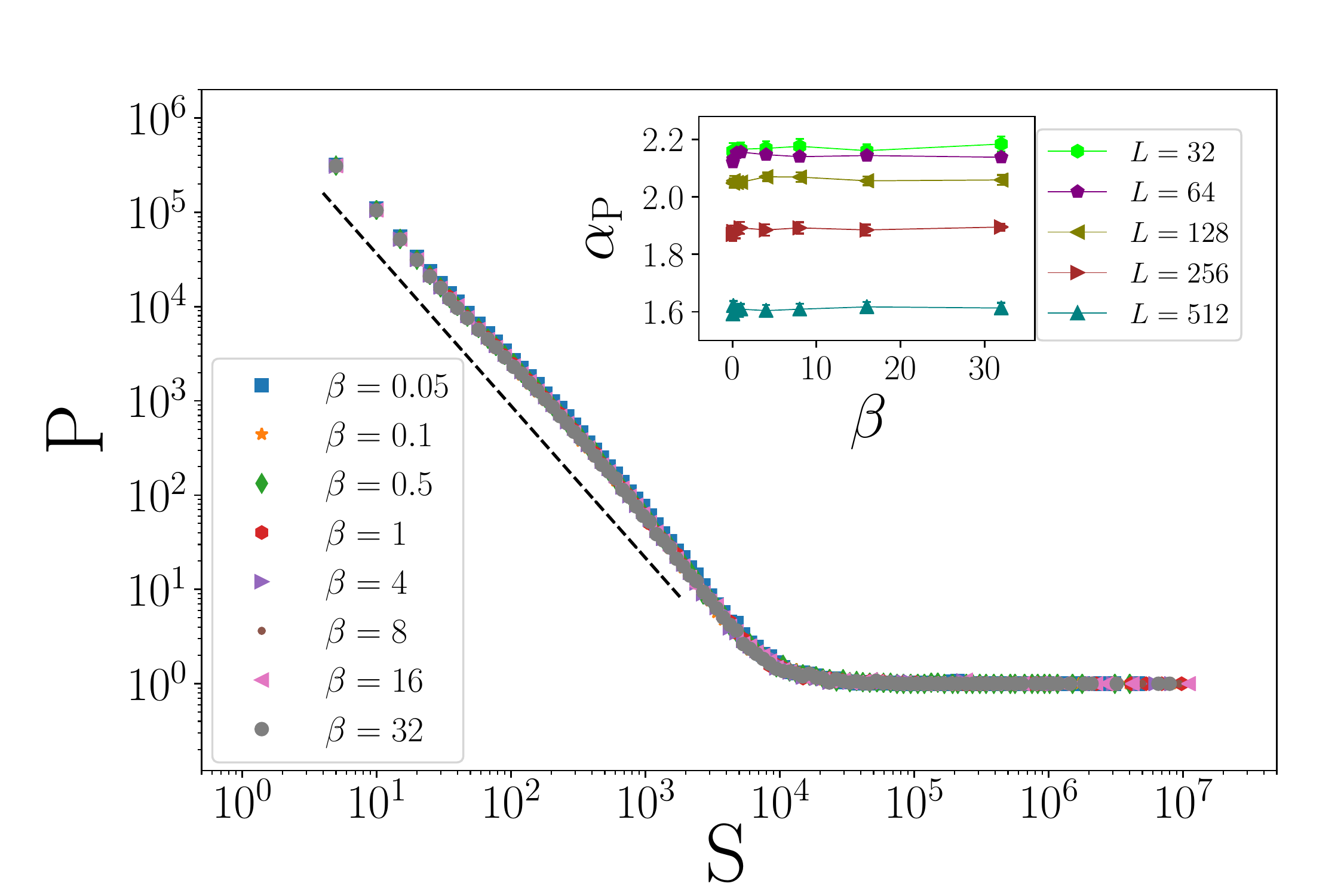}
		\caption{}
		\label{P(s)512winbs.pdf}
	\end{subfigure}
	\begin{subfigure}{0.40\textwidth}\includegraphics[width=\textwidth]{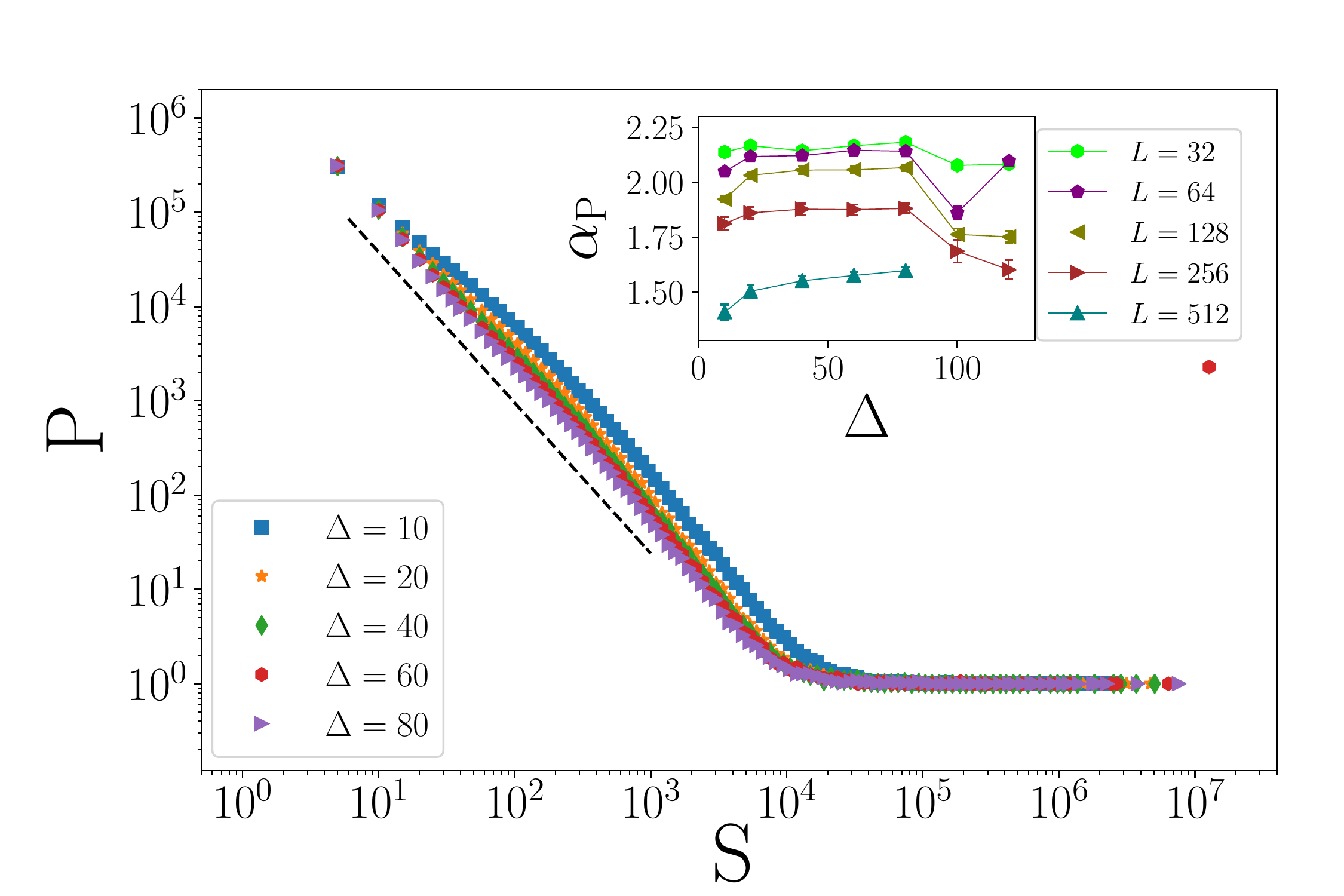}
		\caption{}
		\label{P(s)512winds.pdf}
	\end{subfigure}
	\caption{(Color Online) (a) Distribution function $ P $ in term of size $S$ for various $\beta$ for $L=512$. Inset: ${\alpha_P} $ in term of $\beta$ for various $L$. (b)Distribution function $ P $ in term of time $S$ for various $\Delta$ for $L=512$. Inset: $ {\alpha _P } $ in term of $\Delta$ for various $L$. }
	\label{fig:P512}
\end{figure*}

\begin{figure}
	\begin{subfigure}{0.40\textwidth}\includegraphics[width=\textwidth]{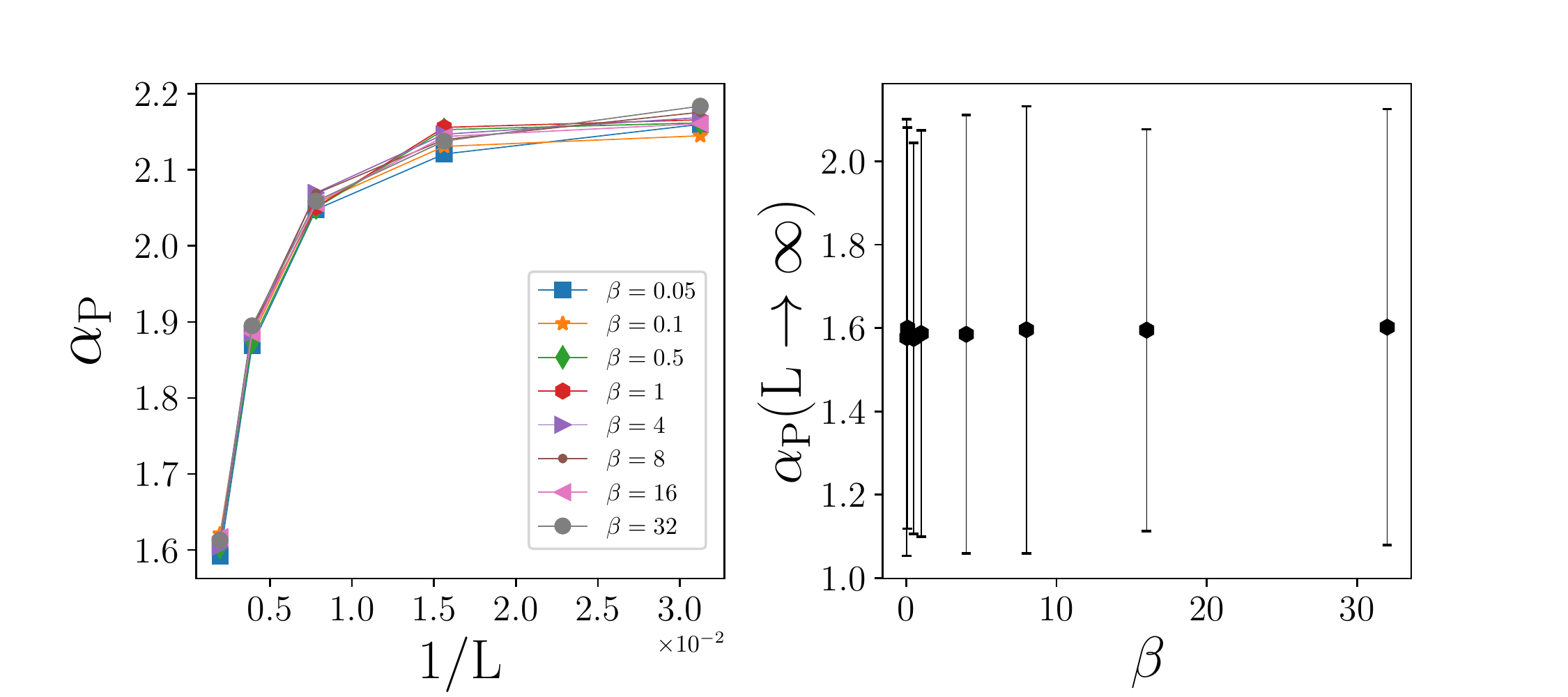}
		\caption{}
		\label{fig:alphaPbs}
	\end{subfigure}
	\begin{subfigure}{0.40\textwidth}\includegraphics[width=\textwidth]{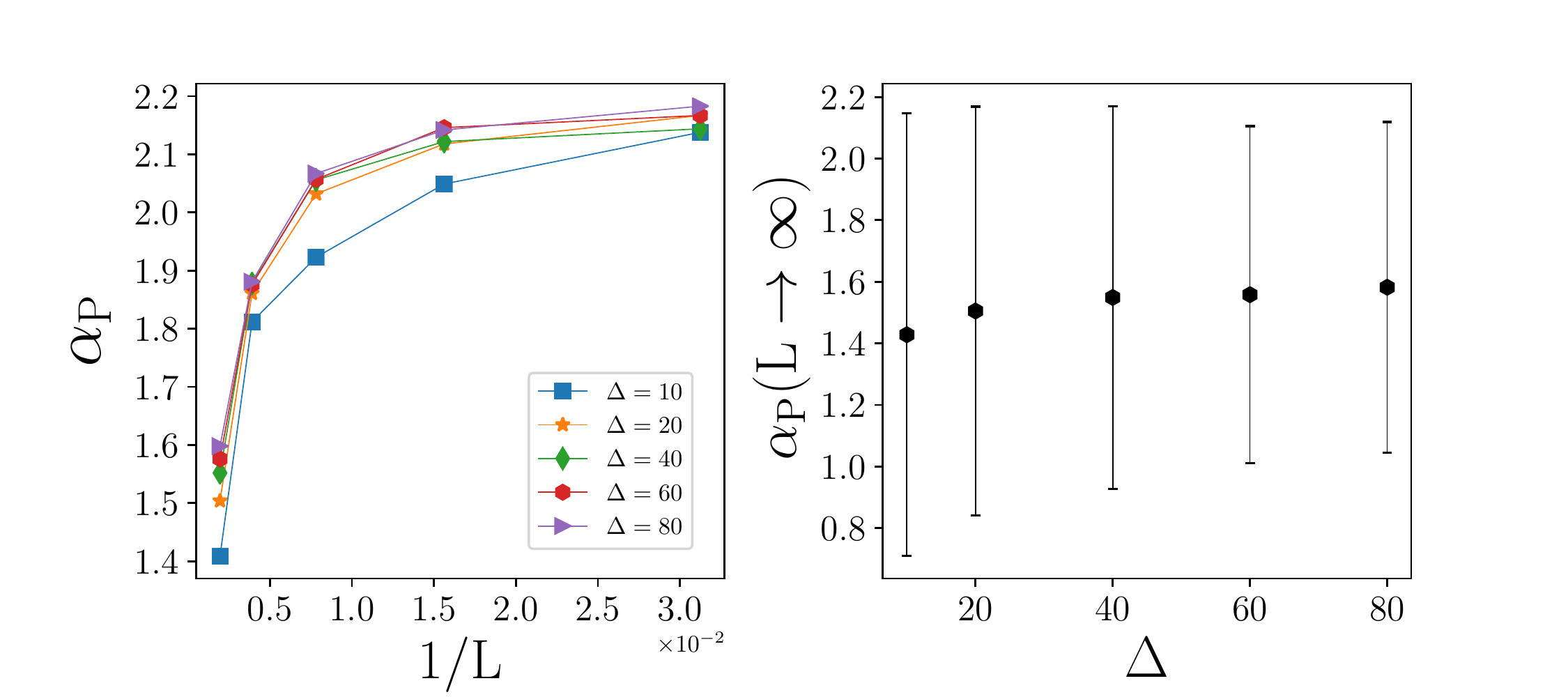}
		\caption{}
		\label{fig:alphaPds}
	\end{subfigure}
	\caption{(a) ${\alpha _P}$ in term of $\frac{1}{L}$ for $L=512$ that fit with $a{x^2} + bx + c$ and report $c$ ($ \alpha_P (L \to \infty $)) in term of $\beta$. (b) ${\alpha _P}$ in term of $\frac{1}{L}$ for $L=512$ that fit with $a{x^2} + bx + c$ and report $c$ ($ \alpha_P(L \to \infty $)) in term of $\Delta$.}
	\label{fig:alphaP}
\end{figure}

\begin{figure*}
	\begin{center}
		\includegraphics[width=100mm]{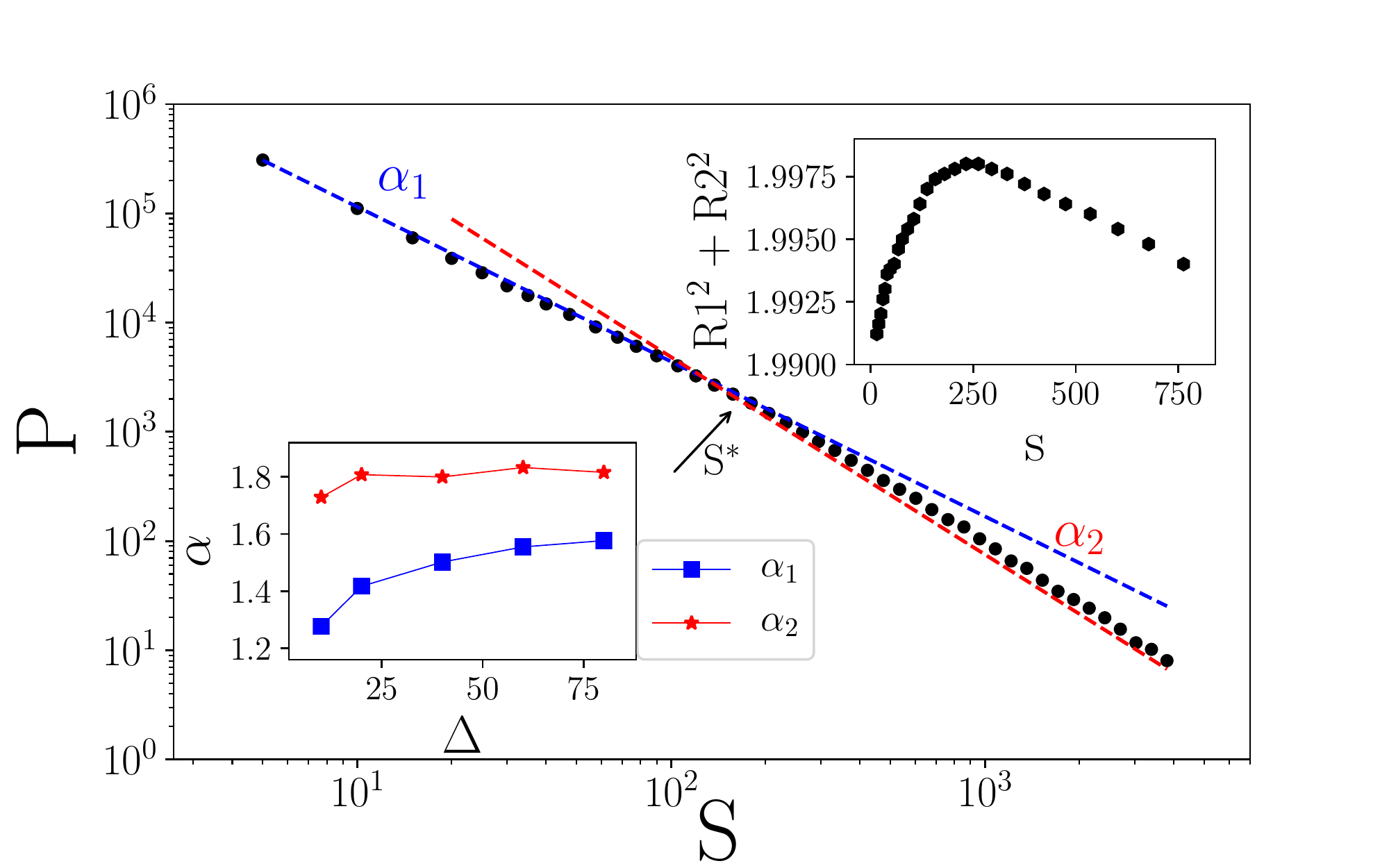}	
	\end{center}
	\caption{ distribution function ( $ P $ ) in term of size ($ S $) for $L=512$ and $\Delta=20$. Up inset: R-square way for findind cross point. Down inset: $\alpha_1$ and $\alpha_2$ for various $\Delta$.}
	\label{fig:Pd60512}
\end{figure*}
We use Monte Carlo Metropolis algorithm explained above. In this paper we prevent the electrons to leave the system from bulk (to the reservoir). We start from a random configuration. As explained above we have two relevant fields: $h$ (from which both the chemical potential and the energy are obtained) and $Z$, which controls the population of the impurities. Algorithmic procedure for obtaining $Z_i$ is to generate a random number $r\in[0,1]$ with uniform distribution, and use the following relation
\begin{equation}
	Z(i)=Z_0+\Delta(r-0.5)
\end{equation}
The exact value of $Z_{0}$ is not essential in simulations since it can easily be absorbed in the chemical potential (considered to be $\mu_{0}=100$ in this paper), so without loose of generality we it to zero. In the primitive stages the system is pretty out of equilibrium and even stationary, i.e. the number of electrons that enter the system is on average larger than the electrons that leave the system. As time goes on, by injecting electrons to random sites through the system, the average electron density in the system increases, until reaching a stationary state after which the average density is almost constant, i.e. the statistical properties of the system doesn't meaningfully change. For instance the number of electrons that enter the system is statistically the same as the number of the electrons that leave the system via the boundaries. We run the code to various amounts of $T$, $\Delta$ and $L$s. The equilibration occurs after each 2000 electron injections, during which $100L^2$ local equilibrations take place. The work function of the boundary sites (from which the electrons can escape) is considered to be $\mu_0$. We consider lattice sizes $L=32,64,128,256,512$ for various amounts of $\Delta$ and $T$. In~\cite{najafi2018percolation} it was shown that a percolation transition occurs in the $\Delta-T$ space. More precisely for each temperature there is a $\Delta^*(T)$ where the system transits from disordered to the ordered phase. \\

In this paper we analyze the time series of the electronic avalanches $S(t)$, $t$ defined as the number of injections. More precisely when $t$th avalanche takes place with non-zero size, then $S(t)$ is the size of this avalanche. Then the the power spectrum is obtained using
\begin{equation}
PS(f) =\lim_{t_{\text{max}}\rightarrow\infty} \frac{1}{t_{\max}}\left|\int_0^{t_{\text{max}}} \text{d}t S(t)\exp \left[ - 2\pi ift \right] \right|^2
\end{equation}
where $f$ is a frequency, and $t_{\max}$ is the maximum time in the analysis. The Auto-correlation function is defined as
\begin{equation}
	AC({t_0}) \equiv \frac{\left\langle {S(t)S(t + t_0)} \right\rangle_T -\left\langle {S(t)} \right\rangle_t^2}{\left\langle {S(t)^2} \right\rangle_T -\left\langle {S(t)} \right\rangle_t^2}
\end{equation}
in which the $t$-average of an arbitrary statistical observable is defined by $ \left\langle O \right\rangle_t \equiv \frac{1}{t_{\max }}\sum\limits_{t = 0}^{t_{\max }} O(S(t))$. For a stationary time series $\left\langle O(t)O(t') \right\rangle$ depends on $t-t'$ (it is invariant under the transformation $t \to t + a$ and $ t' \to t' + a $). For this case the power spectrum is a simple Fourier transformation of the autocorrelation function.

\section{results}
In this section we present the results. Figure~\ref{fig:stationarity} shows the time evolution of the energy of the system ($E_{\beta,\Delta}$), i.e. the process in which the system passes the \textit{transient states}, and reaches a stationary regime at time $t=t_m^{\beta,\Delta}$ which the superscripts $\beta$ and $\Delta$ show that it depends on $\beta$ and $\Delta$. In this figure we re-scaled the time and energy by the relation $\tau\equiv t/t_m^{\beta,\Delta}$ and $\epsilon_{\beta,\Delta}(\tau)\equiv \frac{f_{\beta,\Delta} (\tau ) - 1}{f_{\beta,\Delta} (\tau_m)- 1}$, where $f_{\beta,\Delta}\equiv E_{\beta,\Delta} (\tau )/E_{\beta,\Delta} (0 )$. Figure~\ref{aaveL512betashif.pdf} shows the results for various $\beta$ ($\Delta=1$) and Fig.~\ref{aaveL512deltashif.pdf} it is in terms of $\Delta$ ($\beta=1$). We see that the graphs acceptably fit to each other by this re-scaling, showing that the model shows a universal behavior in terms of $\beta$ and $\Delta$. From the insets we see that $t_m$'s scale with $L^2$ and $f_{\beta,\Delta}(\tau_m)$ saturates in terms of $\beta$, and increases with $\Delta$.\\

The power-spectrum of the time series of the electronic avalanches is analyzed in Fig~\ref{fig:aave512} (in which $\omega$ is the angular frequency) for the maximum size that is considered in this paper, i.e. $L=512$. Figures~\ref{PS(s)512bs.pdf}, which is the power spectrum for various $\beta$ values, and $\Delta=1$ reveals that the power spectrum is power-law for all $\beta$ and $\Delta$ values considered in this work with non-trivial exponent, i.e. $PS(f)\propto f^{-\alpha_{\beta,\Delta}}$ where $\alpha_{\beta,\Delta}$ is the corresponding exponent. In the inset the exponent is shown in terms of $\beta$ and $\Delta$ for various values of $L$. The same analysis has been done in the Fig.~\ref{PS(s)512ds.pdf}, where the points $(\beta_c,\Delta_c)$ on the critical line are considered (see~\cite{Najafi2017Percolation} for the details of the critical line). Although the exponents run with the system size for both cases, they are nearly constant for each size, i.e. they are independent of $\Delta$, signaling the universality of the transition line. The finite size effect is analyzed in Fig~\ref{fig:slopeb512} where the $L$ dependence of the exponents is shown. We found that a second order polynomial function $ax^2+bx+c$ ($x\equiv \frac{1}{L}$) is enough for extrapolation, so that $\alpha_{PS}(L\rightarrow \infty)=c$. Observe that the resulting exponents are in the interval $0.35<\alpha_{PS}<0.65$ in terms of $\beta$ and $0.3<\alpha_{PS}<0.5$ in terms of $\Delta$.\\

To be self-contained, we analyze the autocorrelation function in terms of $\beta$ and $\Delta$. The size dependence of the autocorrelation functions are higher than the power spectrum. To see this, we show these functions for two sizes $L=32$ and $L=512$ in Fig~\ref{fig:Ab32} and Fig~\ref{fig:Ab512}. For the former case some oscillations are observed which exist for all $\beta$ and $\Delta$ values. The autocorrelation functions approach minus values when $t_0\rightarrow 0$, showing the anticorrelation character of the system, i.e. a big (small) avalanche is followed by a small (big) avalanche. The system exhibits an intermittent behavior with time, which gradually gets lost when the system size increases (Fig~\ref{fig:Ab512}). This behavior is also seen on the critical line, i.e. the Fig.~\ref{fig:Ad256}. In fact, the autocorrelation functions are power-law for large enough systems (here $L=512$, Figs.~\ref{fig:A25121} and~\ref{fig:A25122}) which is necessary for having $\frac{1}{f}$ noise. The corresponding exponents (the insets) vary with the systems size ($L=128$, $L=256$ and  $L=512$ here), and we couldn't extract extrapolate the exponents to the thermodynamic limit due to the lack of enough data (note that for smaller sizes the behavior is not power-law). \\

The other important quantity that reflects the state of the system in hand is the distribution function of the size of the electronic avalanches, which has partially been investigated in~\cite{Najafi2017Percolation}. We investigate this quantity along the critical line, and also on the other points (in terms of $\beta$ for $\Delta=1$). Fig~\ref{fig:P512} reveals the power-law structure of this function, showing the scale-invariance of the system. From the Fig~\ref{P(s)512winbs.pdf} we see that the exponents are pretty robust against changing $\beta$, and is size dependent. On the critical line however, it changes a bit and is size-dependent (Fig~\ref{P(s)512winds.pdf}). The extrapolation results (using second order polynomial) are given in Figs.~\ref{fig:alphaPbs} and~\ref{fig:alphaPds} in terms of $\beta$, and along the critical line respectively. The resulting exponent is robust, and fixed on $1.6\pm0.5$ in Fig.~\ref{fig:alphaPbs} ($\Delta=1$, and $\beta$ changes), and is $1.5\pm0.7$ along the critical line.\\

Before closing the paper, it is worth mentioning that the system is not mono-fractal, and the above analysis (for extracting the exponents of the size distribution function) is true only on average. To make this point clear, let us explore the distribution function with more details. Figure~\ref{fig:Pd60512} shows that the distribution function has two different slopes that are separated by $S^*$. The separation point ($S^*$) is identified using $R^2$ test as follows: we fit the first part and second part of the curve which are separated by a (variable) $s_0$ and calculate the corresponding $R_1^2(s_0)$ and $R_2^2(s_0)$. The quantity $r^2(s_0)\equiv R_1^2+R_2^2$ is shown in the inset of Fig.~\ref{fig:Pd60512}, from which we find the true $S^*$ as the point in which $r^2$ becomes maximum. The resulting exponents for the first ($\alpha_1$) and second region ($\alpha_2$) for the critical line is shown in the lower inset, from which we see that $\alpha_2>\alpha_1$ for all $\Delta$ values, both of which being larger the size exponent in the sandpile models.

\section{conclusion}
The $1/f$-noise in two-dimensional electron gas (2DEG) was investigated in this work. The electronic avalanche model which was introduced in~\cite{Najafi2017Percolation} was employed, which is based on the decoherence length in 2DEG and the Thomas-Fermi-Dirac approximation. This model, which involves disorder, was already shown to undergo a percolation phase transition from Metallic to insulating phase at a critical line. We show that this system exhibits flicker noise in all points in the phase space ($T,\Delta$ which are temperature and disorder strength respectively). The exponents of the power spectrum of the system depends weakly on $T$ and $\Delta$ (see Fig.~\ref{fig:slopeb512}), i.e. the exponents are in the interval $0.3<\alpha_{PS}<0.6$. The behavior of the autocorrelation function depend on the system size, i.e. for small system sizes it shows an oscillatory behavior, while for large enough sizes a power-law behavior is observed in terms of time. We analyzed the avalanche size distribution function, and showed that it is also power-law with two exponents, one for small avalanche sizes, and the other for large ones.

\bibliography{refs}

\end{document}